\documentclass[twoside]{article}
\usepackage[a4paper]{geometry}
\usepackage{RR}

\usepackage{graphicx, subfigure}
\graphicspath{{./figures/}}

\usepackage{amssymb}
\usepackage{amsmath}
\usepackage{array}
\usepackage{enumerate}
\usepackage[usenames,dvipsnames]{xcolor}
\usepackage{xspace}

\usepackage{marginnote}
\newcommand{\ans}[1]{#1}

\usepackage{leftidx}

\usepackage[noend]{algpseudocode}
\usepackage{algorithm}

\newlength{\indentation}
\RRetitle{A Robust and Efficient Method for Solving Point Distance Problems by Homotopy}
\RRtitle{Une m\'{e}thode robuste et efficace pour r\'{e}soudre des syst\`{e}mes de contraintes de distances entre points par homotopie}
\titlehead{A Robust and Efficient Method for Solving PDSP by Homotopy }
\RRauthor{
R\'{e}mi Imbach\thanks{Travail en partie effectu\'e au laboratoire ICube.}
\and Pascal Mathis\thanks[IGG]{Laboratoire ICube - UMR 7357 CNRS Universit\'e de Strasbourg, bd S\'ebastien Brant, F-67412 Illkirch Cedex, France}
\and Pascal Schreck\thanksref{IGG}
}

\RRdate{February 2015}
\RRversion{3}
\RRdater{July 2016}

\RRequipe{VEGAS}
\RCNancy

\RRabstract{The goal of Point Distance Solving Problems is to find 2D or 3D placements of points knowing distances between some pairs of points. 
The common guideline is to solve them by a numerical iterative method (\emph{e.g.} Newton-Raphson method). 
A sole solution is obtained whereas many exist. 
However the number of solutions can be exponential and methods should provide solutions close to a sketch drawn by the user.
Geometric reasoning can help to simplify the underlying system of equations by changing a few equations and triangularizing it.
This triangularization is a geometric construction of solutions, called construction plan. We aim at finding several solutions close to the sketch on a one-dimensional path defined by a global parameter-homotopy using a construction plan. Some numerical instabilities may be encountered due to specific geometric configurations. We address this problem by changing on-the-fly the construction plan.
Numerical results show that this hybrid method is efficient and robust.}

\RRresume{Le but de la r\'{e}solution de probl\`{e}mes de contraintes de distances entre points est de placer en 2D ou 3D un ensemble de points connaissant certaines distaces entre paires de points. De tels probl\`{e}mes sont en g\'{e}n\'{e}ral r\'{e}solu gr\^{a}ce \`{a} une m\'{e}thode num\'{e}rique, souvent Newton-Raphson, qui ne produit qu'une solution alors qu'il en existe un nombre exponentiel. Celles ressemblant \`{a} l'esquisse sont d'un int\'{e}ret particulier.
Le raisonnement g\'{e}om\'{e}trique peut cependant aider \`{a} simplifier les syt\`{e}mes d'\'{e}quations correspondant en rempla\c{c}ant quelques \'{e}quations, ce qui permet de les triangulariser. Une telle triangularisation est une construction g\'{e}om\'{e}trique des solutions et est appell\'{e}e plan de construction. On se propose dans ce rapport de trouver plusieurs solutions, proches de l'esquisse sur une courbe d\'efinie par une homotopie utilisant le plan de construction pour r\'{e}duire son co\^{u}t. L'utilisation d'un plan de construction induit des instabilit\'{e}s num\'{e}riques \`{a} proximit\'{e} de certains points; ces instabilit\'{e}s sont \'{e}vit\'{e}s en changeant le plan de construction pendant le suivi de la courbe. La m\'{e}thode d\'{e}crites ici a \'{e}t\'{e} impl\'{e}ment\'{e}e, et les r\'{e}sultats obtenus montrent son efficacit\'{e} et sa robustesse. } 

\RRkeyword{Point Distance Solving Problems, Reparameterization, Curve Tracking, Symbolic-Numeric Algorithm}
\RRmotcle{Probl\`{e}mes de Constraintes de Distances entre Points, Re-param\'{e}trisation, Suivi de courbes, Algorithme symbolique-num\'{e}rique}

\newcommand{\defi}[1]{\emph{#1}}
\newcommand{\domain}[1]{\mathcal{D} {#1}}

\newcommand{\card}[1]{|{#1}|}
\newcommand{\var}[1] {\mathbf{#1}}
\newcommand{\val}[1] {#1}
\newcommand{\OnBranch}[2]{\mathop{}\mathopen{\vphantom{#1}}^{[#1]}\kern-\scriptspace #2}
\newcommand{\ktt}{$K_{3,3}$}
\newcommand{\PosSup}[1]{supp_{+}(#1)}
\newcommand{\Iff}{if and only if}

\newcommand{\qed}{$\Box$}

\newcommand{\R}{\mathbb{R}}
\newcommand{\C}{\mathbb{C}}
\newcommand{\N}{\mathbb{N}}

\newtheorem{defn}{Definition}
\newtheorem{rem}{Remark}
\newtheorem{prop}{Proposition}
\newtheorem{lem}{Lemma}
\newtheorem{thm}{Theorem}
\newtheorem{cor}{Corollary}

\begin{document}
\RRNo{8705}

\makeRR

\section{Introduction}
\label{section_introduction}  

Geometric Constraints Solving Problems arise in many fields such as CAD, robotics or molecular modeling. The problem is to determine the positions of geometric elements (points, lines, planes, circles, etc.) that must satisfy a set of constraints such as distances, angles, tangencies and so on. Commercial solvers generally rely on numerical methods such as Newton or quasi-Newton that provide a single solution. Even when problems are well-constrained the number of solutions may grow exponentially with the number of constraints. Usually the user is not interested in all solutions but only to those whose shape is close to a sketch that he provided. 
 
 Here the restrained class of problems involving only points and distance constraints is considered. Our aim is precisely to design a method that uses the sketch to guide the research of several solutions. We assume that the considered problems are structurally well-constrained. Roughly speaking, this means that there exist some assignments for dimensions leading to finitely many solutions.
 We also consider problems resisting to 
 \ans{decomposition-recombination methods}.

Several methods can yield several or even all the solutions. Subdivision methods \cite{Mourrain2009,Foufou2012} provide all the solutions but the number of boxes to be explored can be huge. In algebraic approaches, homotopy methods have been successfully studied in this area  \cite{durand2000systematic,lamure96solving} but only for small size problems. Indeed, the number of homotopy paths to follow grows exponentially with the number of constraints. 
In \cite{lamure96solving} it is proposed to use the sketch to define a parameter-homotopy. A sole path is followed but a sole solution is obtained.

Another way to get several solutions comes from geometric methods. A construction plan is first derived by applying some geometric construction rules. It consists in  a sequence of basic construction steps. Next, such a plan is numerically evaluated  to yield different solutions. However, no construction plan can be easily found for some 2D problems and for most 3D problems. To circumvent this, \cite{Gao2002} proposes a new approach that performs a \textit{reparameterization}.
In this approach, a geometric constraint system, say $S_1$, is modified by adding and removing some constraints to obtain a system $S_2$ similar to the original one and from which a construction plan can be easily derived. In turn, this construction plan is used to define a reduced system $R$ from the constraints removed from $S_1$. $R$ is then solved by a numerical solver in order to meet the removed constraints while still satisfying the constraints of $S_2$. 
\ans{For point distance problems, the number of constraints that have to be swapped to obtain $S_2$ from $S_1$ is much lower than the number of constraints of $S_1$ hence the size of the system to be numerically solved is drastically reduced.}

The drawback is that the equations of system $R$ are much harder to deal with due to irregular configurations. So the choice of the numerical method is crucial to provide several solutions. In \cite{Gao2002} one or two constraints are removed and the solutions are found by a sampling method. In \cite{FS08}, this idea is extended for more than two constraints, Newton-Raphson method allows to get some of the sought solutions.
In \cite{barki2016re}, the reparameterization is used at a low-level to simplify linear algebra involved by numerical methods.
Finally \cite{imbach2011tracking} presents a first attempt for using a homotopy method along with reparameterization. The idea is to follow a homotopy path to which belongs the sketch. 

All these work can quickly find some of the solutions desired by the user. However some solutions are often missed because of numerical inaccuracies. In this paper we provide an effective and original method to face it. This work is based on 
tracking homotopy paths defined by a construction plan obtained after reparameterization of the problem (system $S_2$). 
The central idea is to detect ill-conditioned configurations induced by the interpretation of the construction plan, 
and to change on-the-fly the construction plan to get away from such configurations.
We justify this approach by showing that these changes of construction plans during paths tracking do not
change the path that is followed.
More precisely,
\begin{itemize}
\item we show that the paths followed with a homotopy method applied to  the reduced system using a construction plan can be glued together around the singular points. The whole path is exactly the path which would be followed by continuation on the original system. Thus, it is independent of the reconstruction plan obtained with the reparameterization phase.
\item we use geometric criteria to detect in advance the singular points caused by a particular construction plan, and we design a way to modify on-the-fly the reparameterization in order to avoid singular points.
\item putting everything together, we marry a homotopy method with a reparameterization to have a new algorithm to solve point distance satisfaction problems. We prove that this algorithm terminates and is correct. We compare the results obtained with our new method with another homotopy method and we find that this is algorithm is 
\ans{more than three}
time faster than our previous algorithm without reparameterization.
\end{itemize}

The rest of the paper is organized as follows. Focusing on 2D problems, 
Sec.~\ref{section_definitions} gives definitions on construction plans and homotopy.  
Sec.~\ref{section_algorithm} gives results about homotopy paths tracking on construction plans that justify our approach.
Sec.~\ref{subsection_guide} explains how to change a construction plan on-the-fly to avoid critical situations. 
The soundness of the approach is justified in Sec.~\ref{section_correctnessTermination}. 
Sec.~\ref{section_generalization} gives tracks to extend our method to 3D problems, and  
Sec.~\ref{section_results} presents some experimental results.

\section{Notations and Definitions}
\label{section_definitions} 

A Point Distance Satisfaction Problem (PDSP) is a constraint satisfaction problem where constraints are imposed distances between points.
Unknown points are sought either in the Euclidean plane for 2D PDSP or in the Euclidean space for 3D PDSP. 
We focus here on the 2D case.

The method presented in this paper uses symbolic manipulations on PDSP to ease their numerical solving.
For the sake of clarity in the description of this symbolic-numeric approach, we will use different 
typefaces to denote a variable and its numeric value.
A boldface lowercase letter as $\var x$ will denote a variable, 
and an uppercase boldface letter as $\var X$ will denote a set of variables.
A value for $\var x$, in general a real number, will be denoted by a lowercase italic letter $\val x$, and a value for $\var X$, in general a real vector, will be denoted by a uppercase italic letter $\val X$.
We make an exception for variables associated with geometric objects: if $\var p$ is a point, we will note $\val p$ a value for $\var p$ whereas it refers to a vector of real values (its coordinates) in a geometric context.  

\subsection{Point Distance Satisfaction Problems}
\label{subsec_defs}

\begin{figure}
\hspace{0.5cm}
\begin{minipage}{0.4\linewidth}
  \begin{small}
\noindent \hspace*{\indentation} \textbf{Unknowns: }\\
\noindent $point$ $\var p_1,..., \var p_6$\\
\noindent \hspace*{\indentation} \textbf{Parameters: }\\
\noindent $length$ $\var a_1,...,\var a_9$\\
\noindent \hspace*{\indentation} \textbf{Constraints: }\\
\noindent $distance(\var p_1,\var p_2)=\var a_1$ \\
\noindent $distance(\var p_2,\var p_3)=\var a_2$ \\
\noindent $...$ \\
\noindent $distance(\var p_2,\var p_5)=\var a_8$ \\
\noindent $distance(\var p_3,\var p_6)=\var a_9$  

  \end{small}
\end{minipage}
\begin{minipage}{0.3\linewidth}
  \centering
  \includegraphics[width=4.5cm]{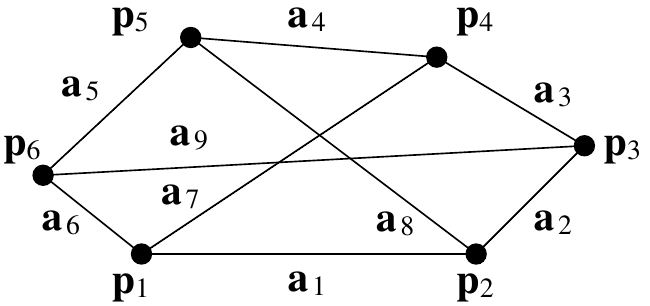}
\end{minipage}

  \caption{A symbolic statement (left part) and a dimensioned sketch (right part) of the PDSP \ktt. 
Edges are distance constraints of parameters $\var a_i$.}
  \label{fig:k33sketch}
\end{figure}

A PDSP $G$ is denoted by $G=C[\var P,\var A]$ where
$\var P$ is a set of unknown points, $\var A$ is a set of length parameters, and $C$ is a set of $m$ constraints of distance.
A distance constraint of parameter $\var a_1 \in \var A$ between points $\var p_2,\var p_3\in \var P$ is written $distance(\var p_2,\var p_3)=\var a_1$. 
A PDSP that consists in constructing $6$ points $\var p_1,\ldots,\var p_6$ in a plane knowing $9$ distances is given in Fig.~\ref{fig:k33sketch}. 
We call it \ktt\footnote{when considering right part of Fig.~\ref{fig:k33sketch} as a non-oriented graph, it is the complete bipartite graph with $3$ vertices in each component}.
Numerical values $\val A^{so}$ for $\var A$ are usually given by a user. The aim is to find the solutions that respect these dimensions.
We suppose in addition that a \defi{sketch}, \emph{i.e.} a geometric placement of points of $\var P$, with possibly a representation of constraints, is available. Right part of Fig.~\ref{fig:k33sketch} shows a sketch of the PDSP \ktt.

Unknown points $\var p_j$ are sought in the Euclidean plane and are each associated with two algebraic unknowns corresponding to their coordinates.
Let $c_i$ be $distance(\var p_j,\var p_k)=\var a_i$. 
It is associated with the numerical function $c_i(\var P,\var A)$ called \defi{numerical interpretation}:
\begin{equation}
\label{eq:dist}
 c_i(\var P,\var A)=\var p_j\var p_k-\var a_i
\end{equation} 
where $\var p_j\var p_k$ holds for the Euclidean distance between $\var p_j$ and $\var p_k$.
Since constraints of distance are invariant up to rigid motions of the plane, placements of points in the plane fulfilling constraints are sought in a \emph{reference}, \emph{i.e.} the values of 3 unknown coordinates are fixed. 
For \ktt~we could search values for points with $\var p_1$ at the origin and $\var p_2$ with null ordinate, and assign variables $\{\var x_1,...,\var x_9\}$ to remaining free coordinates.
 
We denote by $\var X=\{\var x_1,...,\var x_m\}$ the set of free unknown coordinates of points in $\var P$ and we define the system of equations~\ref{eq:SE} associated to $G$ as
\begin{equation}
\label{eq:SE}
 F(\var X,\var A)=0\tag{$\mathcal{F}$} 
\end{equation}
where $F:\R^m\times\R^m\rightarrow\R^m$ has as $i$-th component the numerical interpretation of the constraint $c_i$ defined in Eq.~(\ref{eq:dist}).

We will call \defi{figure} a set of real values $\val X$ for $\var X$. Given positive values $\val A^{so}$ for $\var A$, we call \defi{solution} of $G$  
a figure $\val X$ that is a solution of~\ref{eq:SEParams} defined as
\begin{equation}
\label{eq:SEParams}
 F(\var X,\val A^{so})=0\tag{$\mathcal{F}_{so}$}
\end{equation}

We highlight here that a sketch of $G$ is a figure $X^{sk}$.
In addition, by measuring on $\val X^{sk}$ distances between appropriated points 
one can find positive values $\val A^{sk}$ for $\var A$ such that $\val X^{sk}$ is a solution of the system~\ref{eq:sketchsolution} defined as
\begin{equation}
\label{eq:sketchsolution}
F(\var X,\val A^{sk})=0\tag{$\mathcal{F}_{sk}$}
\end{equation}

\ans{In the following we will consider structurally and generically well constrained PDSP.
A structurally well constrained PDSP satisfies in particular $\card{\var X}=\card{\var A}=\card{C}=m$ if elements of $\var A$ are algebraically independent (see the Koenig-Hall theorem~\cite{plummer1986matching}).
A generically well constrained PDSP admits for generic values of parameters a not null and finite number of solutions. Here generic stands for the complementary of a set having a null Lebesgue measure in an open subset of the space of parameters.}

\subsection{Homotopy}
\label{subsec_homotopy}

Equations of~\ref{eq:SEParams} can be written as polynomials, hence~\ref{eq:SEParams} can be solved in $\C^m$ by a classical homotopy method (see~\cite{allgower1997numerical} for an introduction to homotopy methods, and \cite{durand2000systematic}~for its application to our context). All the complex roots of~\ref{eq:SEParams} are searched and found, 
whereas in applications only the real solutions are relevant.
Here we aim at obtaining 
only
real solutions of~\ref{eq:SEParams} and we use the sketch to define a real homotopy (see~\cite{lamure96solving,imbach2014leading}) between~\ref{eq:sketchsolution} and~\ref{eq:SEParams} using an interpolation of parameters defined as follows. 

\begin{defn}
 \label{defn:interpolationFunction}
 Let $\val A^{sk}=\{a_1^{sk},\ldots,a_m^{sk}\}$ and $\val A^{so}=\{a_1^{so},\ldots,a_m^{so}\}$ be strictly positive real values for $\var A$. 
 We call interpolation function from $\val A^{sk}$ to $\val A^{so}$ a $C^{\infty}$ function $a:\R\rightarrow\R^m$ that satisfies:
 \begin{itemize}
  \item for $1\leq i\leq m, a_i(0)=\val a_i^{sk} \text{ and } a_i(1)=\val a_i^{so}$, and
  \item for $1\leq i\leq m, \forall t\in [0,1],  a_i(t)>0$.
 \end{itemize}
 We call positive support of an interpolation function $a$ and we note it $\PosSup{a}$ the subset of $\R$ where $a_i(t)\geq 0$ for $1\leq i\leq m$.
\end{defn}

Notice that $[0,1]\subset \PosSup{a}$.
Given an interpolation function $a$ from $\val A^{sk}$ to $\val A^{so}$, we define the homotopy system~\ref{eq:homofunctionSPM} as:
\begin{equation}
\label{eq:homofunctionSPM}
 H(\var X,\var t)=F(\var X,a(\var t))=0\tag{$\mathcal{H}$}
\end{equation}
where $H:\R^m\times\R\rightarrow\R^m$ is called the homotopy function associated to $G$.
 
The set of solutions of~\ref{eq:homofunctionSPM} denoted by $H^{-1}(0)$ can be partitioned into a set of connected components that are called \defi{homotopy paths} of~\ref{eq:homofunctionSPM}.
It is worth mentioning here that a point $(\val X, \val t)$ with real components belongs to a homotopy path of~\ref{eq:homofunctionSPM} only if $\val t\in \PosSup{a}$, otherwise two points of $\val X$ would be separated by a negative length.
The following result (see~\cite{imbach2014leading}) underlines the influence of $a$ and $\PosSup{a}$ on the topology of homotopy paths of~\ref{eq:homofunctionSPM}. It is here stated in the general case where constraints are not necessarily distances but also angles, collinearities and so on.

\begin{thm}[\cite{imbach2014leading}]
\label{thm:resultSPM}
Let $H$ define the homotopy~\ref{eq:homofunctionSPM}, $J_H$ be its Jacobian matrix, $\domain{H}\subseteq\R^m\times\R$ be its domain of definition, and $\domain{H}^c$ be its complementary. Under assumptions
 \begin{enumerate}
 \item[(h0)] $\domain{H}$ is open,
 \item[(h1)] $J_H$ has full rank on each point of $H^{-1}(0)$,
 \item[(h2)] $\PosSup{a}$ is compact,
\end{enumerate}
homotopy paths of~\ref{eq:homofunctionSPM} are $1$-dimensional manifolds diffeomorphic either to circles, or to open intervals. 
If an homotopy path $\mathcal{S}$ is diffeomorphic to an open interval then the extremities of $\mathcal{S}$ converge either to a point in $\domain{H}^c$ or to a solution with an infinite norm.
\end{thm}

We make here two remarks to adapt this result in the framework of PDSP. 
First, $H$ is clearly $C^{\infty}$ in $\R^m\times\R$, hence paths cannot converge to a point of $\domain{H}^c$.
Secondly, if $\PosSup{a}$ is compact, $a(\PosSup{a})$ is compact 
and all components of $a(\val t)$ are bounded if $t\in\PosSup{a}$. Hence if $(\val X,\val t)$ belongs to an homotopy path of~\ref{eq:homofunctionSPM}, the components of $\val X$ are coordinates of a set of points lying in a compact. In a PDSP context, Thm.~\ref{thm:resultSPM} can be restated as follows: 

\begin{cor}
 \label{cor:resultSPMadapted}
Let $G$ be a PDSP and $H$ define the homotopy~\ref{eq:homofunctionSPM} satisfying assumptions (h1) and (h2) of Thm.~\ref{thm:resultSPM}. Homotopy paths of~\ref{eq:homofunctionSPM} are $1$-dimensional manifolds diffeomorphic to circles.
\end{cor}

Denoting by $\mathcal{S}$ the homotopy path to which belongs $(\val X^{sk},0)$, $\mathcal{S}$ can be followed with a numerical path tracker until it loops on $(\val X^{sk},0)$. It allows to find points $(\val X,t)$ of $\mathcal{S}$ with $t=1$ that are real solutions of~\ref{eq:SEParams}.  

\subsection{Reparameterization}

A PDSP can be solved very easily when its associated system~\ref{eq:SE} can be organized in a triangular form. 
From a geometric point of view, solving such a system is done by constructing points iteratively as intersections of two circles (three spheres in a 3D context) while making choices between possible intersections. 
The formal statement of the latter geometric construction is called a construction plan.
When a PDSP $G$ cannot be solved with this approach, an idea called reparameterization (see \cite{Gao2002}) is to introduce $d$ new constraints called \defi{added constraints} with unknown parameters called \defi{driving parameters}, in such a way that a construction plan parameterized by driving parameters constructs figures fulfilling all constraints but $d$ that are called \defi{removed constraints}.
$G$ is then solved by finding values of driving parameters such that the figures constructed by the construction plan satisfy the removed constraints.

\subsubsection{Construction Plans}
  \label{subsubsection_CP}
  
\begin{figure}
\hspace{0.5cm}
\begin{minipage}{0.4\linewidth}
  \begin{small}
\noindent \hspace*{\indentation} \textbf{Unknowns: }\\
\noindent $point$ $\var p_1,..., \var p_6$\\
\noindent \hspace*{\indentation} \textbf{Parameters: }\\
\noindent $length$ $\var a_1,...,\var a_8,\var k$\\
\noindent \hspace*{\indentation} \textbf{Constraints: }\\
\noindent $distance(\var p_1,\var p_2)=\var a_1$ \\
\noindent $distance(\var p_2,\var p_3)=\var a_2$ \\
\noindent $distance(\var p_1,\var p_3)=\var k$ \\
\noindent $...$ \\
\noindent $distance(\var p_2,\var p_5)=\var a_8$

  \end{small}
\end{minipage}
\begin{minipage}{0.3\linewidth}
  \centering
  \includegraphics[width=4.5cm]{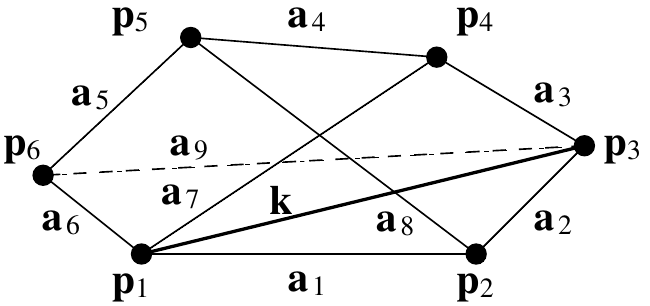}
\end{minipage}

  \caption{A symbolic statement (left) and a dimensioned sketch (right) of the PDSP $q$\ktt.}
  \label{fig:quasik33sketch}
\end{figure}

\begin{figure}
\begin{minipage}{0.4\linewidth}
  \begin{small}
\noindent \hspace*{\indentation} \textbf{Unknowns: }\\
\noindent $point$ $\var p_2,..., \var p_6$\\
\noindent \hspace*{\indentation} \textbf{Parameters: }\\
\noindent $point$ $\var p_1$, $line$ $\var l_1$\\
\noindent $length$ $\var a_1,...,\var a_8,\var k$\\
\noindent \hspace*{\indentation} \textbf{Terms: }\\
\noindent $\var p_2 = InterCL(\var p_1,\var a_1,\var l_1)$ \\
\noindent $\var p_3 = InterCC(\var p_1,\var k,  \var p_2,\var a_2)$ \\
\noindent $\var p_4 = InterCC(\var p_1,\var a_7,\var p_3,\var a_3)$ \\
\noindent $\var p_5 = InterCC(\var p_2,\var a_8,\var p_4,\var a_4)$ \\
\noindent $\var p_6 = InterCC(\var p_5,\var a_5,\var p_1,\var a_6)$ 

  \end{small}
\end{minipage}
\begin{minipage}{0.60\linewidth}
  \centering
  \includegraphics[height=3.7cm]{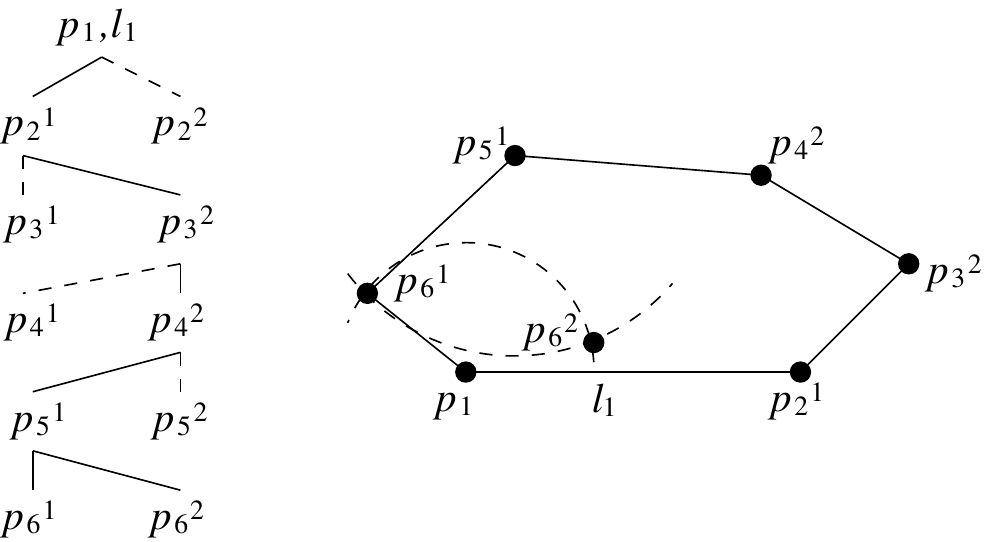}
\end{minipage}

  \caption{Left: a construction plan of solutions of the PDSP $q$\ktt.
  Middle: a sub-tree of the associated interpretation tree. Right: two figures obtained when evaluating the CP on the two branches in solid lines.
}
  \label{fig:quasik33CP}
\end{figure}

Consider the PDSP $q$\ktt~depicted in Fig.~\ref{fig:quasik33sketch} ($q$ holds for ``quasi'') that has been obtained from \ktt~by substituting the constraint $distance(\var p_3, \var p_6)=\var a_9$
by $distance(\var p_1, \var p_3)=\var k$.
Knowing values for $\{\var a_1,\ldots, \var a_8, \var k\}$, its solutions are all found by the simple ruler and compass construction given in the leftmost part of Fig.~\ref{fig:quasik33CP}.
 
The instruction $\var p_2 = InterCL(\var p_1,\var a_1,\var l_1)$ holds for the construction of $\var p_2$ as one of the intersections of the line $\var l_1$ with a circle of center $\var p_1$ and radius $\var a_1$. Here $\var p_1$ and $\var l_1$ are objects of the reference that are fixed to construct solutions up to rigid motions.
The instruction $\var p_i = InterCC(\var p_{i^1},\var a_{i^2},  \var p_{i^3},\var a_{i^4})$ 
holds for the construction of $\var p_i$ as one of the intersections of the circles respectively centered in $\var p_{i^1}$ and $\var p_{i^3}$ of radius $\var a_{i^2}$ and $\var a_{i^4}$.

We will note $I_i[\var p_{i+1},\var A_i]$ the instruction that constructs $\var p_{i+1}$ from objects $\var A_i$ (after a possible re-indexing of points). 
Notice that objects of $\var A_i$ are not only length parameters but also geometric objects of the reference or objects constructed by previous instructions.

\begin{defn}[CP]
A Construction Plan (CP) of objects $\var P$ and parameters $\var A$ with reference $\var A_0$ is a finite sequence $I=(I_i)_{i=1}^{l}$ of terms $I_i[\var p_{i+1},\var A_i]$ in a triangular form, \emph{i.e.}
\begin{enumerate}[(i)]
 \item each $\var p\in \var P$ is either in $\var A_0$ or is constructed by an instruction $I_i\in I$,
 \item for each instruction $I_i[\var p_{i+1},\var A_i]\in I$, each $\var a\in \var A_i$ is either in $\var A_0$, or in $\var A$, or is constructed by a term $I_j\in I$ with $j<i$.
\end{enumerate}
We note it $I[\var P, \var A, \var A_0]$, or more simply $I$. 
\end{defn}

A CP can be seen as a symbolic solution of a set of constraints. 
Consider for instance an instruction $\var p_{i+1} = InterCC(\var p_{i^1},\var a_{i^2},  \var p_{i^3},\var a_{i^4})$, it gives a symbolic solution to the constraints \ans{$distance(\var p_{i+1}, \var p_{i^1})= \var a_{i^2}$} and $distance(\var p_{i+1}, \var p_{i^3})= \var a_{i^4}$.
We associate in such a way a set $C_I$ of constraints with a CP $I$, and we say that 
$I[\var P, \var A, \var A_0]$ is a CP of a PDSP $G=C[\var P, \var A]$ if $C_I=C$.

\subsubsection{Evaluation of a Construction Plan}
  \label{subsubsection_interpretation}
  
Given values $\val A$ and $\val A_0$ for $\var A$ and $\var A_0$, a CP $I[\var P, \var A, \var A_0]$ is evaluated to obtain numerical values of the solutions of constraints $C_I[\var P, \var A]$ by sequentially applying its instructions. At each step a choice between two intersections is done and considering all the possible intersections leads to construct an interpretation tree; its branches bring numerical values for $\var P$. Middle part of Fig. \ref{fig:quasik33CP} shows a sub-tree of the interpretation tree associated to the CP presented in the left part, and in its rightmost part it shows the two figures brought by the two branches in solid line.

Let $I_i[\var p_{i+1},\var A_i]$ be an instruction of $I$. It is interpreted by a multi-function that maps to a value $\val A_i$ of $\var A_i$ the two possible intersection locii of objects defined by $\val A_i$.
We index these locii by an integer and for a given index $b_i$ we note $\OnBranch{\val b_i}{I_i(\val A_i)}$ the function that maps to $\val A_i$ the locus of index $b_i$.
We consider that $\OnBranch{\val b_i}{I_i}$ is not defined when the number of intersections is zero or infinite.
We assume that indexation of intersections is continuous, \emph{i.e} for each $1\leq i\leq l$ and for each index $b_i$, $\OnBranch{\val b_i}{I_i(\val A_i)}$ is $C^{\infty}$ on the interior of its domain of definition.

We call \defi{branch} of $I$ a sequence $b=(b_i)_{i=1}^l$ of indexes and, 
for a given branch $b$, we call \defi{evaluation of $I$ on its branch $b$} the numerical function $\OnBranch{b}{I}(\var A_0,\var A)$ that maps to values $(\val A_0,\val A)$ the composition of functions $\OnBranch{b_i}{I_i}$. 
On a given branch $b$, $\OnBranch{b}{I}$ is $C^{\infty}$ in the interior of its domain of definition as a combination of $C^{\infty}$ functions. 

\subsubsection{Reparameterized Construction Plans}
  \label{subsubsection_RCP}
  
Reparameterized construction plans (RCP) are central objects in the method presented in this paper.
They appear in two steps. First, each problem must be derived in a CP. We are interested in this article in problems whose construction is not known, it is then necessary to transform the problem by adding and removing constraints as explained above. Constraints are added so that a CP is easy to establish.
We do not detail here the way a RCP with a sole driving parameter in each instruction is obtained, see \cite{Gao2002}, \cite{MSI12}, \cite{FS08} for different approaches.  
\ans{This new CP is called a RCP and is completely characterized by: 
the CP itself, the removed constraints that must be satisfied by all solutions and the driving parameters that are the added dimensions.
For instance, a RCP for \ktt~could be given by the CP $I$ given in fig.~\ref{fig:quasik33CP}, the removed constraint $dist(\var p_3,\var p_6)=\var a_9$ and the driving parameter $\var k$.}
Secondly, RCP take also place during the homotopy process. For stability reasons, some distances are removed and others are added on-the-fly according to numerical considerations. 
\ans{The elements of the reference of the CP are also modified during the homotopy process and in the following definition we put forward the reference as one of the characterizing elements of a RCP:}

\begin{defn}[RCP]
 Let $G=C[\var P, \var A]$ a PDSP. A RCP $R$ of $G$ is a quadruplet $(I,C_-,\var A_+,\var A_0)$, where:
 \begin{itemize}
  \item $C_-$ is a subset of $C$ involving parameters $\var A_-$,
  \item $\var A_+$ is a set of parameters called driving parameters,
  \item $I[\var P, \var A', \var A_0]$ is the CP of $C_I[\var P, \var A']$, 
  \item $\var A'= \var A \setminus \var A_- \cup \var A_+$,
  \item $C_I\setminus C$ are distance constraints called added constraints,
 \end{itemize}
and $(I,C_-,\var A_+,\var A_0)$ is such that \ans{$C_- = C \setminus C_I$}.
\end{defn}

In a RCP, the situation where a point is the intersection of two added constraints could not occur. This would create a new point which is not given in the initial statement. 
 So, we will consider RCP that meet the following conditions:
for each circle-circle intersection $\var p_{i+1}=interCC(\var p_{i^1},\var a_{i^2}, \var p_{i^3}, \var a_{i^4})$,
(i) $\var a_{i^2}$ is not a driving parameter (\emph{i.e.} $\var a_{i^2}\in \var A$),
and (ii) if $\var a_{i^4}$ is a driving parameter (\emph{i.e.} $\var a_{i^4}\in \var A_+$), $\var p_{i^3}$ is a reference point (\emph{i.e.} $\var p_{i^3}\in\var A_0$) and is not used as a reference for another instruction.

 Given a RCP with a sole driving parameter in each instruction,
 \ans{it can always be modified to meet the conditions (i) and (ii) as follows:}
 (i) is satisfied by rewriting instructions, and (ii) is satisfied by creating a new reference point for each driving parameter. A RCP for \ktt~satisfying (i) and (ii) is $R'=(I',\{dist(\var p_3,\var p_6)=\var a_9\},\{\var k\},\{\var p_1, \var l_1, \var p_1'\})$
where $I'$ is obtained from $I$ by substituting $\var p_3 = InterCC(\var p_1,\var k,  \var p_2,\var a_2)$ by $\var p_3 = InterCC(  \var p_2,\var a_2,\var p_1',\var k)$.

We focus here on the numerical step of the reparameterization method, that consists in finding values for $\var A_+$ such that figures constructed by $I$ fulfill constraints of $C_-$.

Let $R=(I,C_-,\var A_+,\var A_0)$ be a RCP of $G=C[\var P, \var A]$, $\var A_-$ be the set of parameters of $C_-$ and 
\ans{$\var A'= (\var A \setminus \var A_-) \uplus \var A_+$, where $\uplus$ holds for a disjoint union.
Given a branch $b$, we recall that 
$\OnBranch{b}{I(\var A_0,(\var A \setminus \var A_-)\uplus\var A_+)}$ is the evaluation of $I$ on $b$.
Both for the sake of readability and to make appear the different roles played by driving parameters and other parameters,
we will note it $\OnBranch{b}{I(\var A_0,\var A\uplus\var A_+)}$ even if elements of $\var A_-$ are not involved in $I$.}
When values $\val A_0$ for $\var A_0$ are explicitly fixed, we will note 
$\OnBranch{b}{I(\var A\uplus\var A_+)}$
for 
$\OnBranch{b}{I(\val A_0,\var A\uplus\var A_+)}$.

Let $C_-=\{c_1^-,\ldots,c_d^-\}$. We associate with the RCP $R=(I,C_-,\var A_+,\var A_0)$ the numerical functions
\begin{equation}
\label{eq:RCPfunc}
\begin{array}{cccl}
 \OnBranch{b}{R}:& \R^d\times\R^m &\rightarrow & \R^d \\
   & \var A^+, \var A & \mapsto &
 \left(
\begin{array}{l}
 c_1^-( \OnBranch{b}{I(\var A\uplus\var A_+)}, \var A )\\
 \ldots \\
 c_d^-( \OnBranch{b}{I(\var A\uplus\var A_+)}, \var A )
\end{array}
\right)
\end{array}
\end{equation}
where $b$ is a branch of $I$ and numerical interpretations $c_i^-$ are defined as in Eq.~(\ref{eq:dist}).
Since numerical interpretations $c_i^-$ are $C^{\infty}$, 
functions $\OnBranch{b}{R}$ are $C^{\infty}$ on the interiors of domains of definition of functions $\OnBranch{b}{I}$.

Given values $\val A^{so}$ for $\var A$, there is a one to one correspondence between real solutions of $\OnBranch{b}{R}(\var A^+,\val A^{so})=0$ for all branches $b$ and real solutions of \ref{eq:SEParams}.

\section{Leading Homotopy by Reparameterization}
\label{section_algorithm} 

We aim at finding solutions of~\ref{eq:SEParams} lying on the path $\mathcal{S}$ of~\ref{eq:homofunctionSPM} to which belongs the sketch.
Instead of using a path tracker to follow $\mathcal{S}$ in $\R^m\times\R$, we propose to compute it indirectly 
by following a sequence of paths defined by homotopy functions constructed with a RCP.
These paths are tracked in $\R^d\times\R$ where $d$ is the number of driving parameters of the RCP with $d<<m$, what makes cheaper the path tracking.
We give here a justification of this approach by showing that such paths are diffeomorphic to connected subsets of $\mathcal{S}$.

Sec.~\ref{subsection_assumptions} enumerates assumptions that are required to make our approach valid.
We define in Sec.~\ref{subsection_Rreduced} homotopy functions using RCP and characterize their domains of definition and boundary configurations in Sec.~\ref{subsubsection_boundary}. We establish the link between their paths and 
$\mathcal{S}$ in Sec.~\ref{subsubsection_naive}.

\subsection{Assumptions}
\label{subsection_assumptions}

Let $G=C[\var P, \var A]$ be a PDSP, and $H$ the homotopy function with interpolation function $a$.
Let $R=(I,C_-,\var A_+, \var A_0)$ be a RCP of $G$, where $I=(I_i)_{i=1}^{l}$,
$I_1$ is the instruction $\var p_2=interCL(\var p_1,\var a_1,\var l_1)$ and for $i\geq2$, $I_i$
is the instruction $\var p_{i+1}=interCC(\var p_{i^1},\var a_{i^2},\var p_{i^3},\var a_{i^4})$, 
where $\var a_{i^4}$ is either in $\var A$ or in $\var A_+$. 
Up to a re-indexing, the 
component $a_i$ of $a$ interpolates values of $\var a_i\in \var A$.

The method presented here is valid under the following hypothesis on the interpolation function $a$ and the 
Jacobian matrix $J_H$ of $H$.
\begin{enumerate}
 \item[\textit{(h1)}] $J_H$ has full rank on each point of $H^{-1}(0)$,
 \item[\textit{(h2)}] $\PosSup{a}$ is compact,
 \item[\textit{(h3)}] $\forall i,j$, $a_i(t)=a_j(t)$ has a finite number of solutions,
 \item[\textit{(h4)}] if $t\in \PosSup{a}$, $a_1(\val t)>0$,
 \item[\textit{(h5)}] if $i\geq2$ and $I_i$ is s.t. $a_{i^4}\in\var A$, then $a_{i^2}(\val t)a_{i^4}(\val t)=0 \Rightarrow a_{i^2}(\val t)\neq0 \text{ or } a_{i^4}(\val t)\neq0$ for $t\in \PosSup{a}$,
 \item[\textit{(h6)}] if $i\geq2$ and $I_i$ is s.t. $a_{i^4}\in\var A_+$, then $a_{i^2}(\val t)>0$ for $t\in \PosSup{a}$. 
\end{enumerate}
In what follows, some instructions of the RCP will be changed in such a way that \textit{(h6)} is satisfied only on a subset of $\PosSup{a}$, 
and we will say that \textit{(h6)} is satisfied on a given subset $U$ of $\PosSup{a}$ if \textit{(h6)} holds for each $t$ in $U$.

Let us explain these hypothesis.
\textit{(h1)} and \textit{(h2)} are the hypothesis of Cor.~\ref{cor:resultSPMadapted} and they guarantee that 
paths of \ref{eq:homofunctionSPM} are diffeomorphic to circles. 
\ans{When \textit{(h1)} is satisfied, 
each point of $H^{-1}(0)$ admits a tangent and paths can be tracked with a numerical path-tracker.
Here $H(\var X,\var t)=F(\var X,a(\var t))$ hence \textit{(h1)} 
is strongly related to the rank of $J_F$, the Jacobian matrix of $H$.
Since $G$ is generically well-constrained, $J_F$ has full rank on each point of $F^{-1}(0)$ for generic values of $\var A$ (see Subsec.~\ref{subsec_defs}). 
Verifying that $G$ is generically well-constrained and 
characterizing the interpolation function $a$ to satisfy \textit{(h1)} are both challenging problems and are beyond the scope of this paper.
\cite{li1993solving} justifies real homotopies thanks to the theorem of Sard.
\textit{(h2)} ensures that values $\val t$ such that $H(\var X,\val t)=0$ has real solutions are in a compact interval (see Subsec.~\ref{subsec_homotopy}).
\textit{(h2)} can be satisfied by setting a component $a_i$ of $a$ to $-ct^2 + (a_i^{so} - a_i^{sk} + c)t + a_i^{sk}$
with $c>0$,
and $a_j$ for $j\neq i$ to linear interpolations.
}

Beside its influence on the topology of homotopy paths, $a$ has an impact on the geometric configurations of the figures encountered in such paths.
\ans{Here, we are using RCP to build numerical functions to track homotopy paths of $H$. 
The obtained functions are not defined in the whole space of parameters, and the borders of their domains of definition are characterized in terms of geometric configurations. 
\textit{(h4)}, \textit{(h5)} and \textit{(h6)} restrain the geometrical configurations a path can pass trough.
Configurations that can be encountered are detailed in Subsec.~\ref{subsubsection_boundary}. 
\textit{(h3)} is used to prove the termination of our algorithm.}  

\subsection{R-reduced Homotopy Functions}
\label{subsection_Rreduced}
 
Let $R=(I,C_-,\var A_+,\var A_0)$ be a RCP and suppose values $\val A_0$ for $\var A_0$ are fixed. We call \emph{$R$-reduced homotopy} the homotopies defined by
\begin{equation}
 \label{eq:homotopyRCP}
 \OnBranch{b}{H_R}(\var A^+,\var t)=\OnBranch{b}{R}(\var A^+,a(\var t))=0\tag{$\mathcal{H}_R$}
\end{equation}
where $b$ is a branch of $R$.
The functions $\OnBranch{b}{H_R}:\R^d\times\R\rightarrow\R^d$ are called \emph{$R$-reduced homotopy functions} and
the connected components of solutions of~\ref{eq:homotopyRCP} are called \emph{$R$-reduced paths}.

Considering Eq.~(\ref{eq:RCPfunc}),
the $i$-th component of $\OnBranch{b}{H_R}$ is
$c_i^-( \OnBranch{b}{I(a(\var t)\uplus\var A^+)}, a(\var t) )$.
Noting $\domain{\OnBranch{b}{I}}\subset\R^d\times\R$ the domain of definition of the function defined as
\begin{equation*}
 \ans{\var A^+},\var t \mapsto \OnBranch{b}{I(a(\var t)\uplus\var A^+)}
\end{equation*}
we state that the domain of definition of $\OnBranch{b}{H_R}$ is $\domain{\OnBranch{b}{I}}$, and that $\OnBranch{b}{H_R}$ is $C^{\infty}$ in the interior of $\domain{\OnBranch{b}{I}}$.

We are now interested in characterizing sets $\domain{\OnBranch{b}{I}}$ and their borders in terms of geometric configurations of objects constructed by $I$.
A point $(A^+,t)\in\R^d\times\R$ belongs to the border of $\domain{\OnBranch{b}{I}}$ if its neighborhoods contain points of $\domain{\OnBranch{b}{I}}$ and points where $\OnBranch{b}{I}$ is not defined.
In general, $\domain{\OnBranch{b}{I}}$ is neither open nor close, and 
\ans{contains only a possibly empty subset of}
its border. We will call \defi{boundary} of $\domain{\OnBranch{b}{I}}$ the subset of the border of $\domain{\OnBranch{b}{I}}$ that is in $\domain{\OnBranch{b}{I}}$.

\subsection{Domain of Definition and Boundary Configurations}
\label{subsubsection_boundary}

Since $\OnBranch{b}{I}$ is the combination of functions $\OnBranch{b_i}{I_i}$ we first characterize the domain of definition of the latter functions.

\begin{figure}
  \centering
  \includegraphics[height=2.3cm]{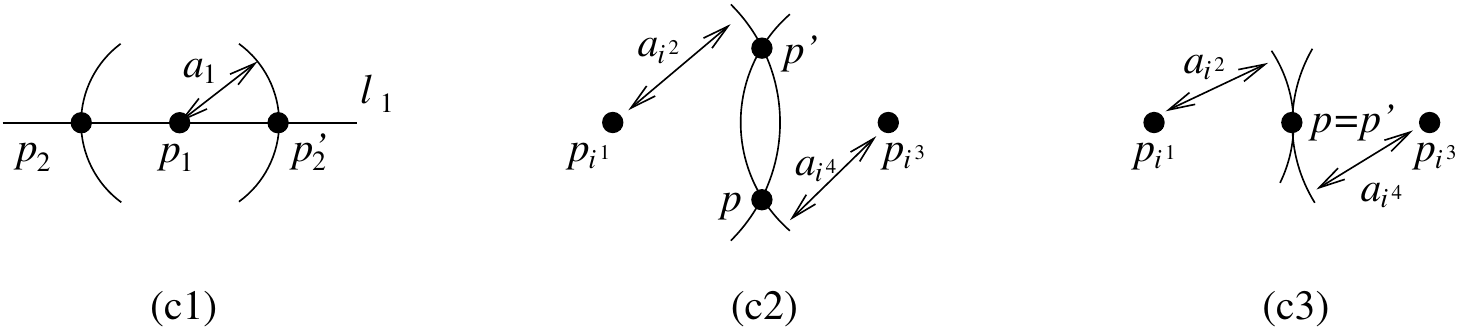}
  \caption{Three geometric configurations \ans{for $InterCL$ and $InterCC$ instructions.}
  }
  \label{fig:boundaryconfigs}
\end{figure}

Let $\var p_2=interCL(\var p_1,\var a_1,\var l_1)$ be the instruction $I_1$.
Since $\var p_1,\var l_1$ are part of the reference their values $\val p_1, \val l_1$ are fixed s.t. $\val p_1\in \val l_1$.
Hence $\OnBranch{b_1}{I_1}$ maps to the value $\val a_1$ of $\var a_1$ one of the intersections $\val p_2, \val p_2'$ of $\val l_1$ with a circle of radius $\val a_1$  which center belongs to $\val l_1$ (see the configuration (c1) on fig.~\ref{fig:boundaryconfigs}).
When $\val a_1>0$, there is an open neighborhood of $\val a_1$ where $\OnBranch{b_1}{I_1}$ is $C^{\infty}$. 
Here we have $\val a_1 = a_1(t)$ and from assumption \textit{(h4)}, $a_1(t)>0$ on $\PosSup{a}$. 

Let $i\geq2$ and $\var p_{i+1}=interCC(\var p_{i^1},\var a_{i^2},\var p_{i^3},\var a_{i^4})$ be the instruction $I_i$ of $I$.
$\OnBranch{b_i}{I_i}$ maps to $(\val p_{i^1},\val a_{i^2},\val p_{i^3},\val a_{i^4})$ one of the two intersections $\val p, \val p'$ of two circles.
We make a disjunction on the number of intersections of the two circles.

When the two circles are disjoint or coincident (with non zero radius), $\OnBranch{b}{I}$ is not defined. 
When the two circles have exactly two intersections, $\OnBranch{b_i}{I_i}$ is clearly $C^{\infty}$ (see the configuration (c2) on fig.~\ref{fig:boundaryconfigs}).

Two configurations can lead the two circles to have exactly one intersection.
The first one is when the latter circles are concentric with null radii, and $\val a_{i^2}=\val a_{i^4}=0$.
Recall that either $\val a_{i^4}\in \var A$ or $\val a_{i^4}\in \var A_+$.
Suppose first $\val a_{i^4}\in \var A$. Hence $\val a_{i^2} = a_{i^2}(t)$ and $\val a_{i^4} = a_{i^4}(t)$, and assumption \textit{(h5)} forbids the situation 
$a_{i^2}(t)=a_{i^4}(t)=0$ when $t\in\PosSup{a}$. 
Suppose now $\val a_{i^4}\in \var A_+$. Hence $\val a_{i^2} = a_{i^2}(t)$ and assumption \textit{(h6)} forbids the situation $a_{i^2}(t)=0$ while $t$ is in a subset for which it holds. 
As a consequence, $\val a_{i^2}=\val a_{i^4}=0$ does not happen
when \textit{(h5)} and \textit{(h6)} hold.

The second configuration is when the two circles are tangent, and at least one circle has a strictly positive radius.
The two centers of circles and their intersection are collinear (see the configuration (c3) on fig.~\ref{fig:boundaryconfigs}).
Clearly, this configuration characterizes the boundary of the domain of definition of the mapping $\OnBranch{b}{I}(\var p_{i^1},\var a_{i^2},\var p_{i^3},\var a_{i^4})$ and is called a \emph{boundary configuration} of $I_i$.

Since $I$ is the combination of its instructions, a point $(\val A_+,\val t)$ is in the boundary of $\domain{\OnBranch{b}{I}}$  only if a boundary configuration holds for $(\val A_+,a(\val t))$ for at least one instruction $I_i$ with $i\geq2$. We will say in this case that the figure $\OnBranch{b}{I(a(t)\uplus\val A_+)}$ presents a boundary configuration of $I$ or $R$, or that $(\val A_+,\val t)$ leads to a boundary configuration of $I$ or $R$.

\subsection{R-Reduced Paths}
\label{subsubsection_naive}

The point here is to characterize $R$-reduced paths, and to link them with paths of~\ref{eq:homofunctionSPM}.
To achieve this, let us define the mappings $\OnBranch{b}{\varphi}:\R^d\times\R\rightarrow\R^m\times\R$ as
\begin{equation}
\label{eq:varphi}
 \OnBranch{b}{\varphi}(\var A_+,\var t) = \left( \begin{array}{c}
                           \OnBranch{b}{I}(a(\var t)\uplus\var A_+) \\
                           \var t
                          \end{array}\right)
\end{equation}
and $\varphi':\R^m\times\R\rightarrow\R^d\times\R$ as
\begin{equation}
\label{eq:varphip}
 \varphi'(\var X,\var t) = \left( \begin{array}{c}
                           c^+_1(\var X) \\
                           \ldots\\
                           c^+_d(\var X)\\
                           \var t
                          \end{array}\right)
\end{equation}
where $c^+_1,\ldots,c^+_d$ are the $d$ added constraints of $R$. $\OnBranch{b}{\varphi}$ are $C^{\infty}$ on $\domain{\OnBranch{b}{I}}$ and $\varphi'$ is $C^{\infty}$ on $\R^m\times\R$.
Consider the following remark, that is a consequence of the characterization of a boundary configuration.
\begin{rem}
 \label{rem:uniqueBranch}
 Let $(\val X, \val t)$ be a point of $H^{-1}(0)$ and $R$ a RCP. $\val X$ does not present any boundary configuration \Iff~it exists a unique branch $b$ and a unique point $(\val A_+, \val t)=\varphi'(\val X, \val t)$ s.t. $(\val X, \val t)=\OnBranch{b}{\varphi}(\val A_+, \val t)$.
 $\val X$ presents a boundary configuration \Iff~there exist at least two branches $b1$ and $b2$ and a unique point $(\val A_+, \val t)=\varphi'(\val X, \val t)$ s.t. $(\val X, \val t)=\OnBranch{b1}{\varphi}(\val A_+, \val t)=\OnBranch{b2}{\varphi}(\val A_+, \val t)$.
\end{rem}

\ans{Let us justify Rem.~\ref{rem:uniqueBranch}.
If $(\val X, \val t)$ is a point of $H^{-1}(0)$, $(\val A_+, \val t)=\varphi'(\val X, \val t)$ is unique by definition of $\varphi'$.
Suppose $X$ does not present any boundary configurations: each point of $X$ is constructed by $R$ by intersecting two circles having two different intersections. If $b1\neq b2$ are two branches of $R$, they correspond to different choices of intersections and $\OnBranch{b1}{\varphi}(\val A_+, \val t)\neq\OnBranch{b2}{\varphi}(\val A_+, \val t)$.
If $X$ has a boundary configuration, at least one point $\var p$ of $\var X$ is the intersection of two circles in configuration (c3). The two branches $b1\neq b2$ corresponding to the same choices for each point but for $\var p$ are such that $(\val X, \val t)=\OnBranch{b1}{\varphi}(\val A_+, \val t)=\OnBranch{b2}{\varphi}(\val A_+, \val t)$.}

We 
show now that $R$-reduced paths are locally diffeomorphic to paths of~\ref{eq:homofunctionSPM}, then we extend the latter diffeomorphism to a global diffeomorphism between pieces of $R$-reduced paths and pieces of paths of~\ref{eq:homofunctionSPM}.

\begin{lem}
\label{lem:RReducedPaths}
 Let $(\val A^+,\val t)$ be s.t. $\OnBranch{b}{H_R}(\val A^+,\val t)=0$, and $\val X=\OnBranch{b}{I}(a(\val t)\uplus\val A^+)$.
 If $(\val A^+, t)$ does not lead to a boundary configuration of $R$, it exists a neighborhood $\mathcal{U}$ of $(\val A^+,\val t)$ and a neighborhood $\mathcal{V}$ of $(\val X,\val t)$ such that $\OnBranch{b}{H_R}^{-1}(0)\cap\mathcal{U}$ is diffeomorphic to $H^{-1}(0)\cap\mathcal{V}$.
\end{lem}
 
\noindent\textbf{Proof of Lem.~\ref{lem:RReducedPaths}:}
Let $\val X=\OnBranch{b}{I}(a(\val t)\uplus\val A^+)$.
Then $\OnBranch{b}{H_R}(\val A^+,\val t)=0\Rightarrow H(\val X,\val t)=0$, and
it exists a homotopy path $\mathcal{S}$ of~\ref{eq:homofunctionSPM} 
to which belongs $(\val X, \val t)$.
Let $\mathcal{H}_-$ be the system of equations having all the equations of~\ref{eq:homofunctionSPM} but the ones corresponding to constraints of $C_-$, and $\mathcal{S}_-$ be the set of its solutions.
$\mathcal{S}_-$ is a $d+1$-dimensional smooth manifold, and $(\val X, \val t)\in\mathcal{S}_-$ and $\mathcal{S}\subseteq\mathcal{S}_-$ hold. 
As a consequence, $\mathcal{S}$ is a $1$-dimensional smooth submanifold of $\mathcal{S}_-$, and in any open neighborhood of $(\val X,\val t)$ in $\mathcal{S}_-$, $\mathcal{S}$ is a $1$-dimensional smooth manifold.

Since $(\val A^+, t)$ does not lead to a boundary configuration, $\OnBranch{b}{\varphi}$ is $C^{\infty}$ on
an open neighborhood $\mathcal{U}\subset\R^d\times\R$ of $(\val A_+,\val t)$. Let $\mathcal{V}=\OnBranch{b}{\varphi}(\mathcal{U})$. Clearly $\mathcal{V}\subset\mathcal{S}_-$ and $(\val X,\val t)\in\mathcal{V}$ hold.
We show that $\mathcal{V}$ is an open neighborhood of $(\val X,\val t)$ in $\mathcal{S}_-$.
Let $\varphi'_{\mid \mathcal{S}_-}:\mathcal{S}_-\rightarrow\R^d\times\R$ be the restriction of $\varphi'$ to $\mathcal{S}_-$.
$\varphi'$ is $C^{\infty}$ on $\mathcal{S}_-$ and $\varphi'(\mathcal{V})=\mathcal{U}$. 
Hence $\mathcal{V}$ is an open neighborhood of $(\val X,\val t)$ in $\mathcal{S}_-$ as the inverse image of an open neighborhood.

We finish the proof by remarking that the mapping $\OnBranch{b}{\varphi}_{|\mathcal{U}}:\mathcal{U}\rightarrow\mathcal{V}$, which inverse is $\varphi'_{\mid \mathcal{S}_-}$, is a diffeomorphism.
\qed

The following Prop. extends the property of Lem.~\ref{lem:RReducedPaths} to a global property.

\begin{prop}
 \label{prop:link}
 Let $\mathcal{S}$ be a homotopy path of \ref{eq:homofunctionSPM} and $R$ be a RCP.
 Let $\mathcal{S}'\subseteq\mathcal{S}$ be a connected subset of $\mathcal{S}$ that does not contain any figure with a boundary configuration of $R$. Then it exists a unique branch $b$, a unique $R$-reduced path $\OnBranch{b}{\mathcal{S}}$ and a 
 subset $\OnBranch{b}{\mathcal{S}}'\subseteq\OnBranch{b}{\mathcal{S}}$ such that $\OnBranch{b}{\mathcal{S}}'$ is diffeomorphic to $\mathcal{S}'$ by $\OnBranch{b}{\varphi}$. 
\end{prop}

\noindent\textbf{Proof of Prop.~\ref{prop:link}:}
Let $(X,t)$ be a point of $\mathcal{S}'$. Since $\val X$ does not present a boundary configuration of $R$, it exists (see Rem.~\ref{rem:uniqueBranch}) a unique branch $b$ and 
a unique point $(\val A_+, \val t)=\varphi'(\val X, \val t)$ such that $\OnBranch{b}{\varphi}(\val A_+, \val t)=(X,t)$.
Hence $\OnBranch{b}{H_R}(\val A_+, \val t)=0$ and $(\val A_+, \val t)$ belongs to an homotopy path $\OnBranch{b}{\mathcal{S}}$ of $\OnBranch{b}{H_R}^{-1}(0)$.

Since $\mathcal{S}'$ is connected and $\varphi'$ is $C^{\infty}$, $\varphi'(\mathcal{S}')$ is a connected subset of $\R^d\times\R$.
Moreover, $\varphi'(\mathcal{S}')$ belongs to the interior of $\domain{\OnBranch{b}{I}}$ 
otherwise a point of $\varphi'(\mathcal{S}')$ would belong to the boundary of $\domain{\OnBranch{b}{I}}$ and would lead to a boundary configuration of $R$.
As a consequence, $\OnBranch{b}{\varphi}$ is well defined on $\varphi'(\mathcal{S}')$.

We show now that $\varphi'(\mathcal{S}')\subseteq\OnBranch{b}{\mathcal{S}}$. If it is not the case, it exists at least another 
branch $b2$ and another $R$-reduced
path $\OnBranch{b2}{\mathcal{S}}$ with $\varphi'(\mathcal{S}')\subseteq(\OnBranch{b}{\mathcal{S}}\cup\OnBranch{b2}{\mathcal{S}})$, and $\OnBranch{b}{\mathcal{S}}\cap\OnBranch{b2}{\mathcal{S}}\neq\emptyset$ holds from the local diffeomorphism property. Let $(\val A_+', \val t')\in (\OnBranch{b}{\mathcal{S}}\cap\OnBranch{b2}{\mathcal{S}})$. Then $(\val A_+', \val t')$ is such that 
$(\val X', \val t')=\OnBranch{b}{\varphi}(\val A_+',\val t')=\OnBranch{b2}{\varphi}(\val A_+', \val t')$ and $(\val X',\val t')\in\mathcal{S}'$ presents a boundary configuration of $R$ from Rem.~\ref{rem:uniqueBranch}. 

Let us defined $\OnBranch{b}{\mathcal{S}}'$ as $\varphi'(\mathcal{S}')$. It is a connected subset of $\OnBranch{b}{\mathcal{S}}$, and $\OnBranch{b}{\varphi}(\OnBranch{b}{\mathcal{S}}')=\mathcal{S}'$. 
Finally $\OnBranch{b}{\varphi}$ is injective and is a global diffeomorphism from $\OnBranch{b}{\mathcal{S}}'$ to $\mathcal{S}'$. 
\qed

Prop.~\ref{prop:link} states that it is possible to compute a part of $\mathcal{S}$ that does not contain any boundary configurations by following a path of $\OnBranch{b}{H_R}^{-1}(0)$.

Now, assuming that the figures of $\mathcal{S}$ presenting a boundary configuration of $R$ are in a finite number 
(boundary configurations of type (c3) can be each described by a polynomial equation, hence figures presenting a boundary configuration of $R$ can be seen as the solutions of systems of $m+1$ polynomials involving $m+1$ variables)
the set $\mathcal{S}$ can be written $\mathcal{S}^1\cup (\val X^1,\val t^1)\cup \mathcal{S}^2\cup \ldots \cup\mathcal{S}^n\cup \val (\val X^n,\val t^n)$, where $(\val X^i,\val t^i)$ are such that $\val X^i$ presents a boundary configuration of $I$, and $\mathcal{S}^i\subseteq\mathcal{S}$ are connected, pairwise disjoint and does not contain any figure presenting a boundary configuration.
From Prop.~\ref{prop:link} each path $\mathcal{S}^i$ can be computed by following a path $\OnBranch{bi}{\mathcal{S}}$ of $\OnBranch{bi}{H_R}^{-1}(0)$, hence $\mathcal{S}$ can be computed by following the sequence of paths $\OnBranch{bi}{\mathcal{S}}$ and making the appropriate branch changing in points $(\val X^i, \val t^i)$.

\begin{figure}
  \begin{minipage}{0.5\linewidth}
  \centering
  \includegraphics[height=5cm]{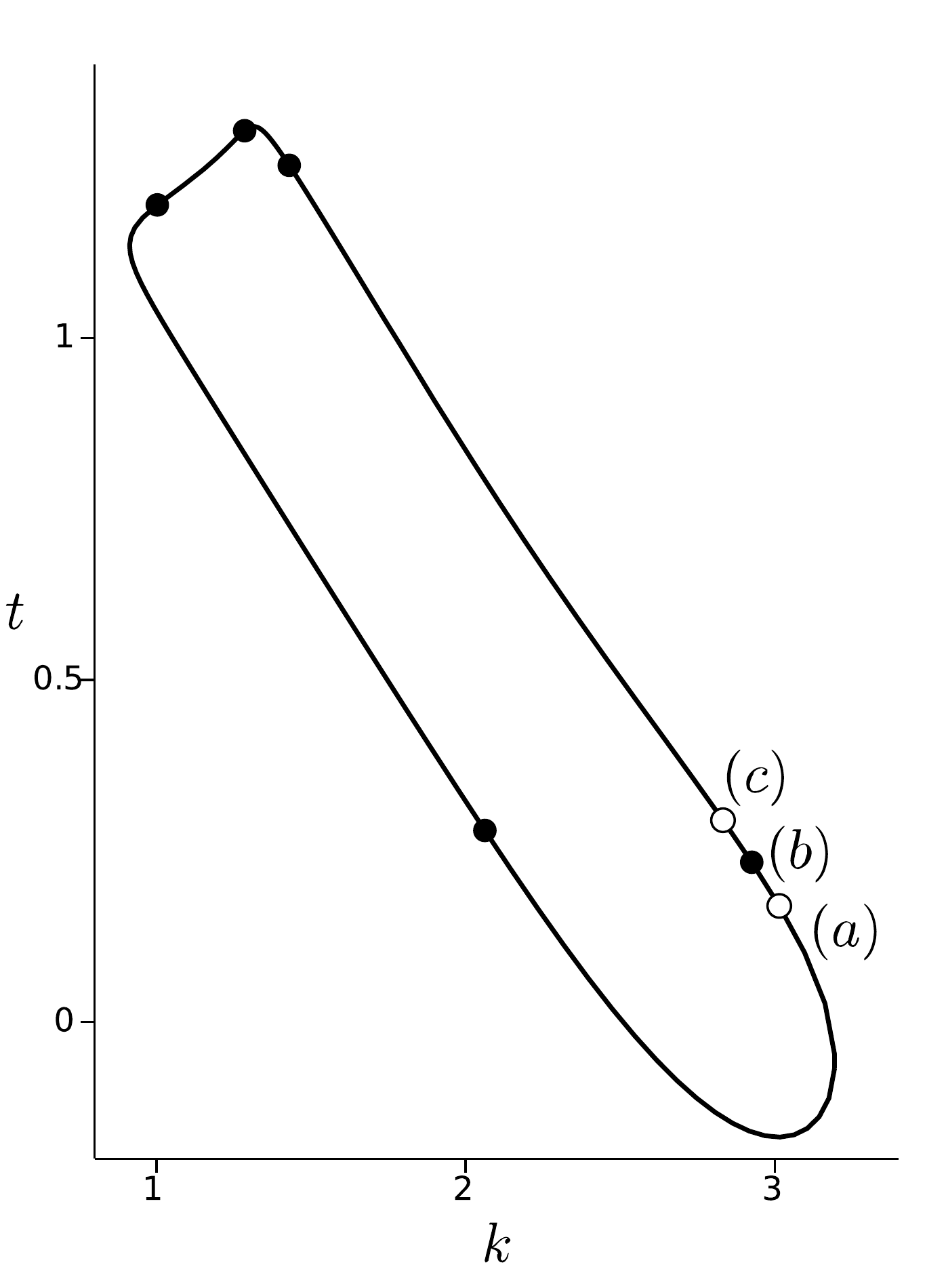}
  \end{minipage}
  \begin{minipage}{0.5\linewidth}
   \centering
  \includegraphics[height=5cm]{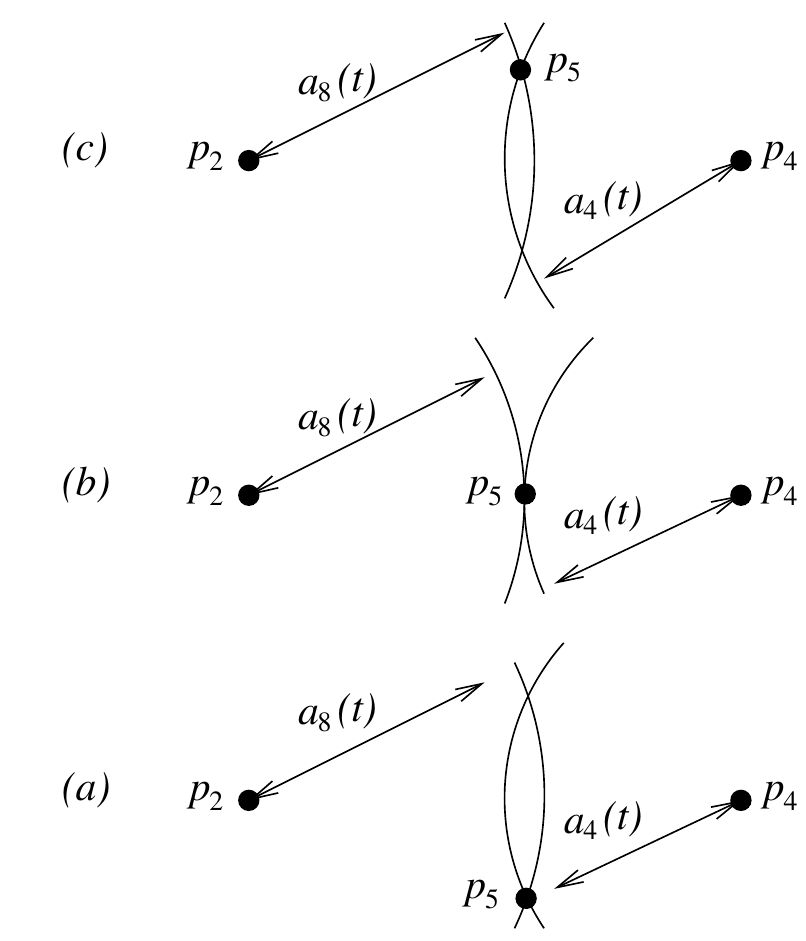}
  \end{minipage}
  \caption{Left: the sequence of $R$-reduced paths corresponding to a path $\mathcal{S}$ of~\ref{eq:homofunctionSPM} for \ktt. 
   Black circles mark points leading to a a boundary configuration.
   Right: geometric configurations corresponding to points $(a)$, $(b)$, $(c)$.}
   \label{fig:pathsreparamK33}
\end{figure}

The left part of fig.~\ref{fig:pathsreparamK33} shows the sequence of $R$-reduced paths $\OnBranch{bi}{\mathcal{S}}$ corresponding to a path of~\ref{eq:homofunctionSPM} for the PDSP \ktt.
Its right part shows geometric configurations near a point leading to a boundary configuration.

\section{On-The-Fly Change of the RCP}
\label{subsection_guide}

The method roughly depicted above hides a pitfall:
it leads to follow $R$-reduced paths until a boundary configuration is reached.
But $R$-reduced paths are numerically bad behaving near boundary configurations.

It can be easily seen by considering a function that maps to a positive real number $\val k$ the positive $y$-coordinate of the intersections of the unit circle with a circle centered in $(0,2)$ of radius equal to $k$. Noting $y$ this function, we have $y(\val k)=\dfrac{\sqrt{-\val k^4+10\val k^2-9}}{4}$ and $y'(\val k)=\dfrac{-4\val k^3+20\val k}{8\sqrt{-\val k^4+10\val k^2-9}}$.
The domain of definition of $y$ is $[1,3]$ which boundaries $1,3$ corresponds to configurations where the circles are tangent; we have $lim_{\val k\rightarrow\{1,3\}}y'(\val k) =+\infty$, and it is not due to the chosen system of coordinates. 

Such unbounded values of derivatives highly affect the efficiency of a numerical path tracking, that proceeds by approximating a path by its tangent.

We propose here to introduce a measure of the distance from a figure  of $\mathcal{S}$, or from a point of a $R$-reduced path, to a boundary configuration, and to
stop the tracking process of a $R$-reduced path when this distance is smaller than a real parameter $\alpha$.
Then we change either the RCP or the values of its references in a way that the distance to a boundary configuration is greater than $\alpha$. The new $R$-reduced homotopy path is then followed to compute the path $\mathcal{S}$.
This process is repeated each time the distance to a boundary configuration is smaller than $\alpha$. 

The distance to a boundary configuration is defined in Sec.~\ref{subsubsection_distance}, and our algorithm to change on-the-fly the RCP is described in Sec.~\ref{subsecton_on-the-fly}. We first give an intuition of our approach on the example \ktt~in Sec.~\ref{subsec_overview}.

\subsection{Overview of our Method on an Example}
\label{subsec_overview}

\begin{figure}
  \begin{minipage}{0.34\linewidth}
  \centering
  \includegraphics[height=5cm]{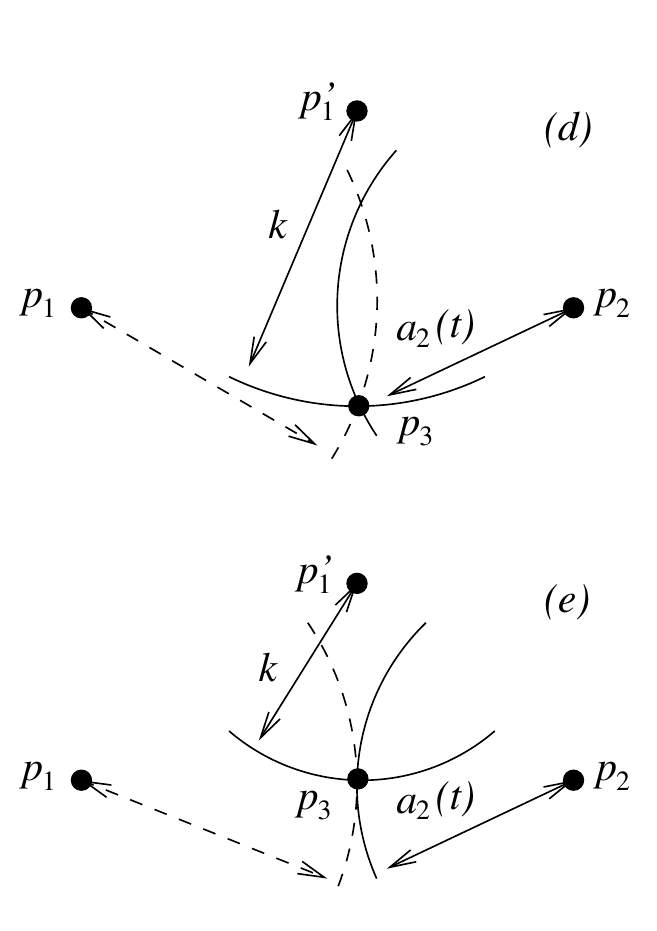}
  \end{minipage}
  \begin{minipage}{0.3\linewidth}
  \centering
  \includegraphics[height=5cm]{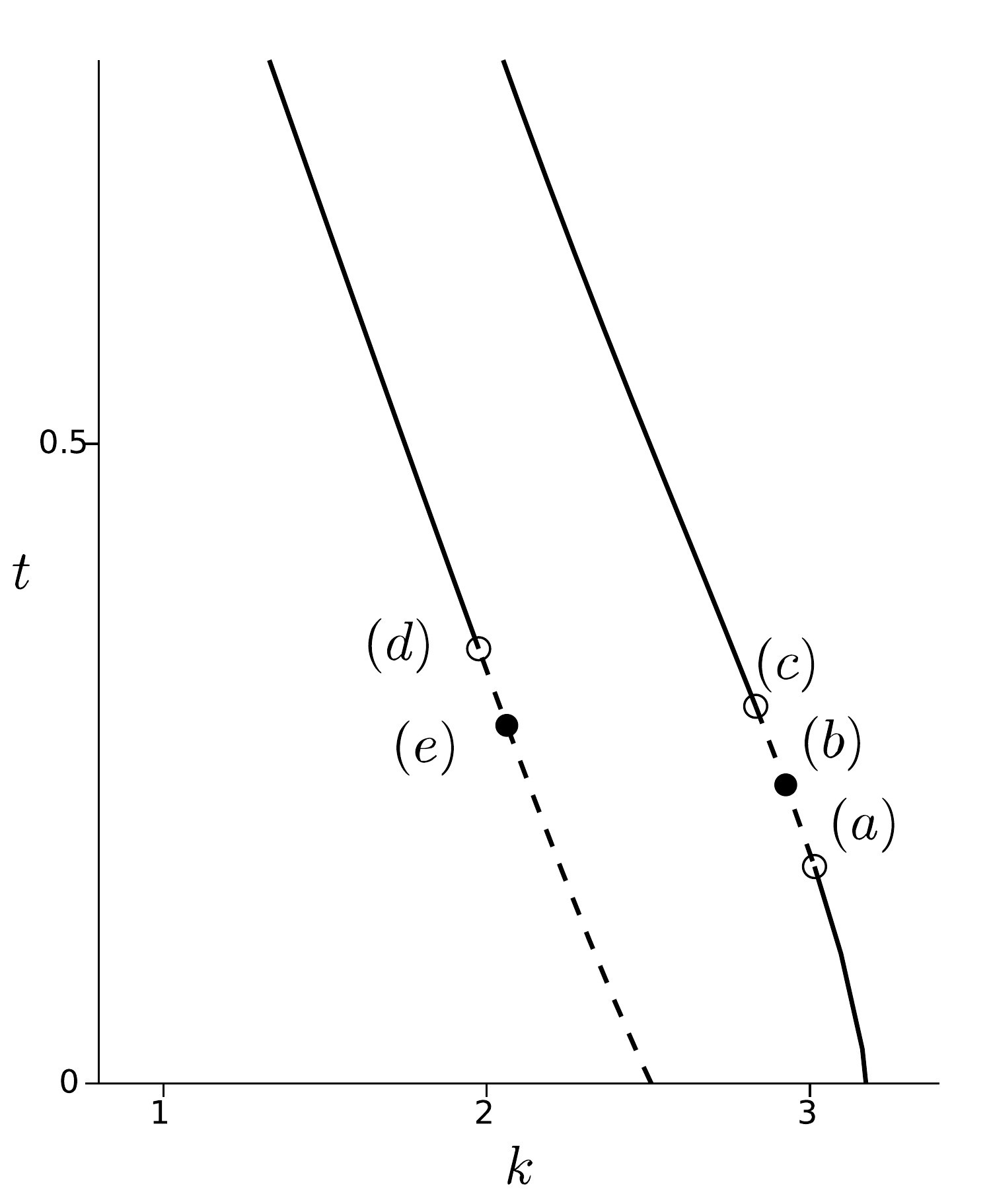}
  \end{minipage}
  \begin{minipage}{0.34\linewidth}
   \centering
  \includegraphics[height=5cm]{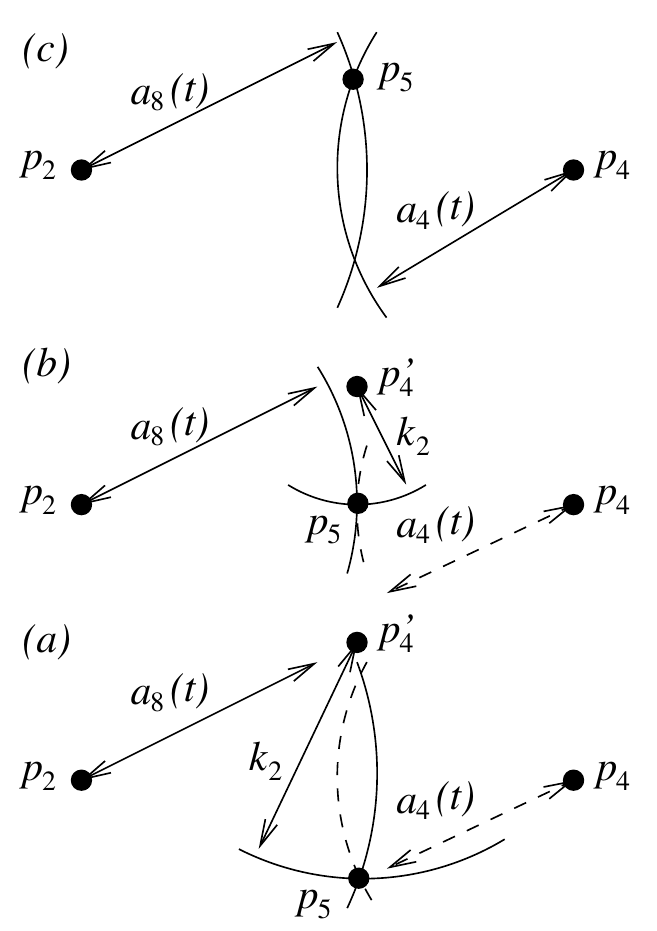}
  \end{minipage}
  \caption{Center: a zoomed view of a sequence of $R$-reduced paths for \ktt. Pieces in solid lines are followed by our algorithm.
           Black circles mark points leading to a a boundary configuration.
           Left and Right: geometric configurations corresponding to points $(a),\ldots (d)$. Dashed double arrows correspond to constraints removed by our algorithm.}
   \label{fig:pathsreparamK33Algo}
\end{figure}

The middle part of Fig.~\ref{fig:pathsreparamK33Algo} shows a zoomed view of pieces of $R$-reduced paths ($R$ is given in Sec.~\ref{subsubsection_RCP}) for~\ktt.
When following this union of paths from the sketch, the point $(a)$ is first reached.
The points $\val p_5,\val p_4,\val p_2$ in the figure $X^{a}$ constructed in $(a)$ is shown in the right part of Fig.~\ref{fig:pathsreparamK33Algo}.
The two circles constructed when evaluating the instruction $\var p_5 = interCC(\var p_2, \var a_8, \var p_4, \var a_4)$ are almost tangent, and $X^{a}$ is ``too close'' to a boundary configuration
(this notion will be detailed in Sec.~\ref{subsubsection_distance}).

To avoid the point $(b)$ that leads to a boundary configuration, a new driving parameter and a new reference point is added to the RCP: the instruction $\var p_5 = interCC(\var p_2, \var a_8, \var p_4, \var a_4)$ is replaced by $\var p_5 = interCC(\var p_2, \var a_8, \var p'_4, \var k_2)$, where $\var p'_4$ is a new reference point and $\var k_2$ is a new driving parameter.
The constraint $distance(\var p_4, \var p_5)=\var a_4$ is added to the set of removed constraints to guarantee that it is fulfilled by constructed figures.
For an appropriated placement of $\var p'_4$, presented in Sec.~\ref{subsubsection_shifting}, 
the figure constructed by the CP is not too close to a boundary configuration. The new RCP defines new paths, that can be followed while avoiding the boundary configuration of point $(b)$ (see right part of Fig.~\ref{fig:pathsreparamK33Algo}).
When reaching the point $(c)$, the original instruction $\var p_5 = interCC(\var p_2, \var a_8, \var p_4, \var a_4)$ can be restored while staying ``far away'' from a boundary configuration.
The piece of path between $(a)$ and $(c)$ in the middle part of Fig.~\ref{fig:pathsreparamK33Algo} is drawn in dashed line to underline that it is a projection of the path that is followed with our algorithm. 

Suppose now that the point $(d)$ (see central part of Fig.~\ref{fig:pathsreparamK33Algo}) is reached. The constructed figure 
is too close to a boundary configuration of the instruction $\var p_3 = interCC(\var p_2, \var a_2, \var p_1', \var k)$ where $\var k$ is a driving parameter. In that case, the point $\var p_1'$ is moved in order to avoid the boundary configuration of point $(e)$ (see left part of Fig.~\ref{fig:pathsreparamK33Algo}). The piece of path after $(d)$ is drawn in dashed line to figure out that 
it is no longer the path that is followed by our algorithm.

\subsection{Distance to a Boundary Configuration}
\label{subsubsection_distance}

Let $I_i$ be $\var p_{i+1}=interCC(\var p_{i^1},\var a_{i^2}, \var p_{i^3}, \var a_{i^4})$. We associate with $I_i$ the real function $\gamma_i$
taking its values in $[0,1]$:
\begin{equation*}
\label{gammaInterCC}
 \ans{\gamma_i(\var X) = \dfrac{dist(\var p_{i+1}, (\var p_{i^1},\var p_{i^3}) )}{max(\var p_{i+1}\var p_{i^1},\var p_{i+1}\var p_{i^3})}}
\end{equation*}
where $dist(\var p_{i+1}, (\var p_{i^1}, \var p_{i^3}) )$ is the distance between $\var p_{i+1}$ and the line $(\var p_{i^1},\var p_{i^3})$. 

$\gamma_i$ is defined and continuous on $\val X$ when points $\{\val p_i,\val p_{i^1},\val p_{i^3}\}\subset\val X$ are not coincident.
Notice it never happens
when $I_i$ is an instruction of a RCP for which assumptions \textit{(h5)} and \textit{(h6)} hold.
$\gamma_i$ vanishes only on figures presenting a boundary configuration of $I_i$.

We associate to $(I_i)_{i=1}^l$ the function $\gamma_I$ that measures the distance of a figure $X$ to a boundary configuration of $I$ defined as 
\begin{equation*}
\label{gamma}
 \gamma_I(\var X) := min_{2\leq i\leq l}\gamma_i(\var X).
\end{equation*}
$\gamma_I$ is defined and continuous at least when assumptions \textit{(h5)} and \textit{(h6)} hold, 
and vanishes on figures presenting a boundary configuration of $I$.
In the following, we will note $\OnBranch{b}{\gamma_I}(\var A_+, \var t)$ for $\gamma_I(\OnBranch{b}{I}(a(\var t)\uplus\var A_+))$.

\subsection{Path Tracking with On-The-Fly Change of the RCP.}
\label{subsecton_on-the-fly}

\begin{algorithm}[t]
    \caption{Path tracking with RCP swapping}
\label{algo:main-algorithm}
\begin{algorithmic}[1]
 \Require{ A RCP $R=(I,C_-,\var A_+,\var A_0)$,
           an interpolation function $a$, a sketch $\val X^{sk}$, $\alpha$.}
 \Ensure{a list of solutions $\mathcal{L}_{sol}$}
 
 \State Let $\val t=0$, $\val X = \val X^{sk}$
 \State Find $\val A_+, \val A_0, b$ s.t. $\OnBranch{b}{I}(\val A_0,a(t)\uplus\val A_+) = \val X$, fix $\var A_0$ to $\val A_0$
 \While{True}
   \State Follow the path $\OnBranch{b}{\mathcal{S}}\subset\OnBranch{b}{H_R}^{-1}(0)$ from $(\val A_+,t)$ {\bf while checking}
   \If{$\OnBranch{b}{\mathcal{S}}$ passes trough the sketch}
     \State \Return $\mathcal{L}_{sol}$
   \ElsIf{$\OnBranch{b}{\mathcal{S}}$  passes trough hyperplane $\var t=1$}
     \State Append current figure to $\mathcal{L}_{sol}$
     \State Set $(\val A_+,t)$ to current point
   \ElsIf{current point $(\val A^{cur}_+,\val t^{cur})\in\OnBranch{b}{\mathcal{S}}$ satisfies (sc1) {\bf or} $\OnBranch{b}{\gamma_I}(\val A^{cur}_+,\val t^{cur})\leq\alpha$}
     \State Let $\val X = \OnBranch{b}{I}(a(\val t^{cur})\uplus\val A^{cur}_+)$
     \State Apply Algo.~\ref{algo:change-RCP} to change $R=(I,C_-,\var A_+,\var A_0)$, and obtain $\val A_{+}$, $\val A_0$
     \State Find $b$ s.t. $\OnBranch{b}{I}(\val A_0,a(t)\uplus\val A_+) = \val X$, fix $\var A_0$ to $\val A_0$
   \EndIf
 \EndWhile
\end{algorithmic}
\end{algorithm}
\floatname{algorithm}{Algorithm}

Algo.~\ref{algo:main-algorithm} describes the main process of our method. 
We consider a PDSP $G=C[\var X, \var A]$, a RCP $R$, a sketch $\val X^{sk}$ and an interpolation function $a$ from 
$\val A^{sk}$ to $\val A^{so}$.
We assume that assumptions \textit{(h1)} to \textit{(h6)} are satisfied for $a$ and $R$.

In the step 2, value $\val A_0$ for elements of $\var A_0$ 
are read on $\val X$ and fixed. Values $\val A_+$ for $\var A_+$ are obtained by evaluating $\varphi'(\val X, \val t)$.
In steps 2 and 13, the branch $b$ is found by evaluating one by one instructions $I_i$ on each branch, and keeping for each $b_i$ the choice that leads to construct $\val X$. 

When entering in the step 4, a point $(\val X,\val t)$ of the path $\mathcal{S}$ of~\ref{eq:homofunctionSPM} to which belongs the sketch is known as well as a point $(\val A_+,\val t)$ with $\OnBranch{b}{\varphi}(\val A_+,\val t) = (\val X,\val t)$.
We temporary assume \textit{(h7)}: $\val X$ is not closer than $\alpha$ to a boundary configuration.  

In the step 4, an abstract path-tracker is used to follow in a given orientation the $R$-reduced path $\OnBranch{b}{\mathcal{S}}$ s.t. $(\val A_+,\val t)\in\OnBranch{b}{\mathcal{S}}$. We assume that it allows to compute in a finite 
number of iterations
the connected subset $\OnBranch{b}{\mathcal{S}}'\subseteq\OnBranch{b}{\mathcal{S}}$ s.t. points of $\OnBranch{b}{\mathcal{S}}'$ are not closer than $\alpha$ to a boundary configuration, and that it stops when a point at a distance $\alpha$ to a boundary configuration is reached 
or when the stopping condition (sc1) described below is satisfied.
It is also assumed that it is possible to detect when $\OnBranch{b}{\mathcal{S}}'$ passes trough hyperplanes $\var t=0$ and $\var t=1$, and to get exact intersections with latter hyperplanes.

The tracking process stops when a point $(\val A_+,\val t)$ of $\OnBranch{b}{\mathcal{S}}$ is at a distance $\alpha$ to a boundary configuration. In this case a new RCP, or at least new values $\val A_0, \val A_+$ for reference points and driving parameters, is computed thanks to Algo.~\ref{algo:change-RCP} described in Sec~\ref{subsubsection_changing}, and the process re-enters in step 4 with a new RCP $R$, new values $\val A_0, \val A_+$ and a new branch $b$ such that $(\val A_+, \val t)$ is not closer than $\alpha$ to a boundary configuration. 
Hence assumption \textit{(h7)} is satisfied when re-entering step 4. If \textit{(h7)} is not satisfied when performing for the first time step 4, Algo.~\ref{algo:change-RCP} is directly applied.

Stopping condition (sc1) is detailed in Sec.~\ref{subsubsection_stopping}. It is satisfied when it is not possible to ensure that assumption \textit{(h6)} holds. When (sc1) is satisfied, Algo.~\ref{algo:change-RCP} changes the RCP in such a way \textit{(h6)} holds on the computed path.  

Sought solutions are found when $\OnBranch{b}{\mathcal{S}}$ passes through the hyperplane $\var t = 1$, and the overall process is stopped when a
point $(\val A_+,\val t)$ of $\OnBranch{b}{\mathcal{S}}$ is s.t. $\OnBranch{b}{I}(a(\val t)\uplus\val A_+)=\val X^{sk}$. Latter termination criterion is checked each time the the hyperplane $\var t = 0$ is crossed.

The orientation used to follow the new $R$-reduced path is chosen in order to avoid backtrack on $\mathcal{S}$. We will discuss practical details of path-tracking in Sec.~\ref{subsection_trackingImp}.
We now focus on the description of the way the RCP is changed.

\subsubsection{Changing RCP}
\label{subsubsection_changing}

\begin{algorithm}[t]
    \caption{Change RCP}
\label{algo:change-RCP}
\begin{algorithmic}[1]
 \Require{A RCP $(I,C_-,\var A_+,\var A_0)$, 
          a table $T$ of original instructions,
          a current figure $X$,
          $\alpha$}
 \Ensure{New values $\val A_+,\val A_0$}
 
 \For{$2\leq i \leq l$}
  \State Let $I_i\in I$ be $\var p_{i+1}=interCC(\var p_{i^1},\var a_{i^2},\var p_{i^3},\var a_{i^4})$
  \If{$\gamma_i(\val X)\leq \alpha$ {\bf and} $T[i]==\emptyset$}
    \If{$\var a_{i^4}\in \var A_+$}
     \State Apply Algo.~\ref{algo:shift-reference} to obtain $\val a_{i^4},\val p_{i^3}$
    \Else \Comment{assume $\val a_{i^2}\geq\val a_{i^4}$}
     \State Let $\var p_{i^3}'$ be point and $\var a_{d+1}$ a length parameter 
     \State Let $\val p_{i^3}'=\val p_{i^3}$ and $\val a_{d+1}=\val a_{i^4}$
     \State Let $I_i'$ be $\var p_{i+1}=interCC(\var p_{i^1},\var a_{i^2},\var p_{i^3}',\var a_{d+1})$
     \State Apply Algo.~\ref{algo:swap-instruction} with $I_i$ and $I_i'$ as inputs 
     \State Apply Algo.~\ref{algo:shift-reference} to obtain $\val a_{d+1}, \val p_{i^3}'$
    \EndIf
  \ElsIf{{\bf not} $T[i]==\emptyset$}
   \State Let $I_i'=T[i]$ associated with $\gamma_i'$
   \State Let $I_i'$ be $\var p_{i+1}=interCC(\var p_{i^1},\var a_{i^2},\var p_{i^3}',\var a_{i^4}')$ 
   \If{$\gamma_i'(\val X)> \alpha$} \Comment{restore $I_i'$}
    \State Apply Algo.~\ref{algo:swap-instruction} with $I_i$ and $I_i'$ as inputs
   \ElsIf{$\val a_{i^2}<\val a_{i^4}'$} \Comment{change driving parameter}
     \State Apply Algo.~\ref{algo:swap-instruction} with $I_i$ and $I_i'$ as inputs
     \State {\bf goto} step 3
    \ElsIf{$\gamma_i(\val X)\leq \alpha$} \Comment{move point}
      \State Apply Algo.~\ref{algo:shift-reference} to obtain $\val a_{i^4},\val p_{i^3}$
   \EndIf
  \EndIf
 \State Actualize $\val A_+,\val A_0$ and $d=\card{\var A_+}$
 \EndFor
 \State \Return $\val A_+,\val A_0$
\end{algorithmic}
\end{algorithm}

In Algo~\ref{algo:main-algorithm}, when a point $(\val A_+, \val t)$ s.t. $\OnBranch{b}{\gamma_I}(\val A_+, \val t)=\alpha$ is reached, the RCP or at least the values of the reference are changed. 
The basic principle of the mechanism that changes the RCP or the reference is to identify the instruction(s) $I_i$ s.t. $\gamma_i(\val X)= \alpha$, where $X=\OnBranch{b}{I}(a(\val t)\uplus\val A_+)$.

Let $I_i$ be $\var p_{i+1}=interCC(\var p_{i^1},\var a_{i^2}, \var p_{i^3}, \var a_{i^4})$ s.t. $\gamma_i(\val X)= \alpha$. 
Then either $\var a_{i^4}\in\var A_+$ and $\var p_{i^3}\in\var A_0$ is a reference point involved only in $I_i$,
or $\var a_{i^4}\in \var A$, $\var p_{i^3}\notin \var A_0$ and we suppose without loss of generality that $\val a_{i^2}\geq\val a_{i^4}$ (otherwise arguments of the instruction are swapped).

In the first case, new values for $\var p_{i^3}, \var a_{i^4}$ are computed s.t. $\gamma_i(\val X)> \alpha$ thanks to Algo.~\ref{algo:shift-reference}.
In the second case, $I_i$ is exchanged with the instruction $I_i'$ defined as $\var p_{i+1}=interCC(\var p_{i^1},\var a_{i^2}, \var p_{i^3}', \var a_{d+1})$
and involving the new driving parameter $\var a_{d+1}$ and the new reference point $\var p_{i^3}'$.
Then values $\val a_{d+1}, \val p_{i^3}'$ are computed s.t. $\gamma_i(\val X)> \alpha$, where $\gamma_i$ is the distance to a boundary configuration of $I_i'$. 
This relaxation is counterbalanced by adding $distance(\var p_{i+1},\var p_{i^3})=\var a_{i^4}$ to the set $C_-$ of removed constraints of the new RCP.

A table $T$, that does not appear in Algo.~\ref{algo:main-algorithm} to ease its description, is used to save the instructions of the original RCP (\emph{i.e.} given as input of Algo.~\ref{algo:main-algorithm}). 
Entries of $T$ are initially empty, and each time an instruction $I_i$ of the original RCP is exchanged with $I_i'$, $I_i$ is stored in the $i$-th entry of $T$.
When $I_i$ could be restored while ensuring that the distance to a boundary configuration stays greater than $\alpha$, $I_i$ is restored and $T[i]$ is re-set to $\emptyset$. 
Hence an instruction $I_i$ of the current RCP is an instruction of the original RCP if $T[i]$ is empty.

\begin{algorithm}[t]
    \caption{Swap instructions}
\label{algo:swap-instruction}
\begin{algorithmic}[1]
 \Require{A RCP $R=(I,C_-,\var A_+,\var A_0)$, an index $i$,
          a new instruction $I_i'$, a table $T$ of original instructions}
 
 \State Let $I_i\in I$ be $\var p_{i+1}=interCC(\var p_{i^1},\var a_{i^2},\var p_{i^3},\var a_{i^4})$
 \State Let $I_i'$ be $\var p_{i+1}=interCC(\var p_{i^1},\var a_{i^2},\var p_{i^3}',\var a_{i^4}')$
 \State Let $c$ be the constraint $distance(\var p_{i+1},\var p_{i^3})=\var a_{i^4}$
 \If{{\bf not} $c\in C_-$} \Comment{introduce new instruction}
  \State $\var A_+\leftarrow\var A_+\cup \{\var a_{i^4}'\}$
  \State $\var A_0\leftarrow\var A_0\cup\{\var p_{i^3}'\}$
  \State $I\leftarrow(I_{1}\cup\ldots\cup I_{i-1} \cup I_i' \cup I_{i+1}\cup\ldots\cup I_l)$
  \State $C_-\leftarrow C_-\cup\{c\}$
  \State $T[i] \leftarrow I_i$
 \Else \Comment{restore original instruction}
  \State $\var A_+\leftarrow\var A_+\setminus \{\var a_{i^4}'\}$
  \State $\var A_0\leftarrow\var A_0\setminus\{\var p_{i^3}'\}$
  \State $I\leftarrow(I_{1}\cup\ldots\cup I_{i-1} \cup I_i' \cup I_{i+1}\cup\ldots\cup I_l)$
  \State $C_-\leftarrow C_-\setminus\{c\}$
  \State $T[i] \leftarrow \emptyset$
 \EndIf
 \State \Return
\end{algorithmic}
\end{algorithm}

Algo.~\ref{algo:change-RCP} details the mechanism to change a RCP, and Algo.~\ref{algo:swap-instruction} details the way instructions are swapped.
They both modify in place the RCP. 
Notice that it could exist several indices $i$ such that $\gamma_i(X)=\alpha$, and that Algo.~\ref{algo:change-RCP} is designed to take it into account.

\subsubsection{Stopping Condition (sc1)}
\label{subsubsection_stopping}
The tracking process in Algo.~\ref{algo:main-algorithm} also stops when the condition (sc1) is satisfied for a current point $(\val A_+^{cur},\val t^{cur})$. We define here this stopping condition.

Let $I_i$ be an instruction of $I$ s.t. $T[i]\neq\emptyset$. 
Hence $I_i$ has been introduced by Algo.~\ref{algo:change-RCP} to replace the original instruction $I_i'=T[i]$.
Let $I_i$ be $\var p_{i+1}=interCC(\var p_{i^1},\var a_{i^2},\var p_{i^3}',\var a_{i^4}')$, and $I_i'$ be $\var p_{i+1}=interCC(\var p_{i^1},\var a_{i^2},\var p_{i^3},\var a_{i^4})$.
Recall that in our homotopy context, values for $\var a_{i^2}, \var a_{i^4}$ are $\val a_{i^4}(t)$ and $\val a_{i^2}(t)$.

(sc1) is satisfied if it exists $2\leq i \leq l$ s.t. $T[i]\neq\emptyset$ and $a_{i^4}(t^{cur})>a_{i^2}(t^{cur})$.
When (sc1) is satisfied on a point of a path, Algo.~\ref{algo:change-RCP} is called. 
Unless the instruction $I_i$ making (sc1) to be satisfied has been restored in steps 15-16,
the if condition in step 17 of Algo.~\ref{algo:change-RCP} is satisfied, and steps 18-19 are performed: the original instruction $I_i'$ is restored, and when entering step 6, its arguments are swapped
(\emph{i.e.} $\var p_{i+1}=interCC(\var p_{i^1},\var a_{i^2},\var p_{i^3},\var a_{i^4})$ is replaced by $\var p_{i+1}=interCC(\var p_{i^3},\var a_{i^4},\var p_{i^1},\var a_{i^2})$). Then a new driving parameter is introduced.

When returning to Algo.~\ref{algo:main-algorithm} after Algo.~\ref{algo:change-RCP} have been performed, the stopping condition (sc1) is not satisfied.
We will state in Sec.~\ref{subsection_pointi} that this mechanism ensures that \textit{(h6)} holds.

\subsubsection{Shifting Reference}
\label{subsubsection_shifting}

\begin{algorithm}[t]
    \caption{Shift Reference }
\label{algo:shift-reference}
\begin{algorithmic}[1]
 \Require{An instruction $\var p_{i+1}=interCC(\var p_{i^1},\var a_{i^2}, \var p_{i^3}, \var a_{i^4})$, 
          values $\val p_{i+1},\val p_{i^1},\val a_{i^2},\val p_{i^3},\val a_{i^4}$ for $\var p_{i+1},\var p_{i^1},\var a_{i^2},\var p_{i^3},\var a_{i^4}$.}
 \Ensure{New values $\val p_{i^3}', \val a_{i^4}'$ for $\var p_{i^3}, \var a_{i^4}$}
 
 \State Let $\val p$ be the projection of $\val p_{i+1}$ on the line $(\val p_{i^1},\val p_{i^3})$
 \State Let $\val p_{i^3}'=\val p_{i+1} + \val a_{i^2}\frac{\overrightarrow{\val p_{i+1} \val p}}{\val p_{i+1}\val p}$ 
 \State Let $\val a_{i^4}' = \val a_{i^2}$
 \State \Return $\val p_{i^3}'$, $\val a_{i^4}'$
\end{algorithmic}
\end{algorithm}

Algo.~\ref{algo:shift-reference} is called in steps 5, 11 and 21 of Algo.~\ref{algo:change-RCP}. 
It computes values for driving parameters and reference point of an instruction $I_i$ in order that the constructed figure $\val X$ satisfies $\gamma_i(X)>\alpha$.
The following proposition states that this goal is achieved after applying Algo.~\ref{algo:shift-reference} if $0<\alpha<\frac{1}{2}$.

\begin{figure}
  \centering
  \includegraphics[width=5cm]{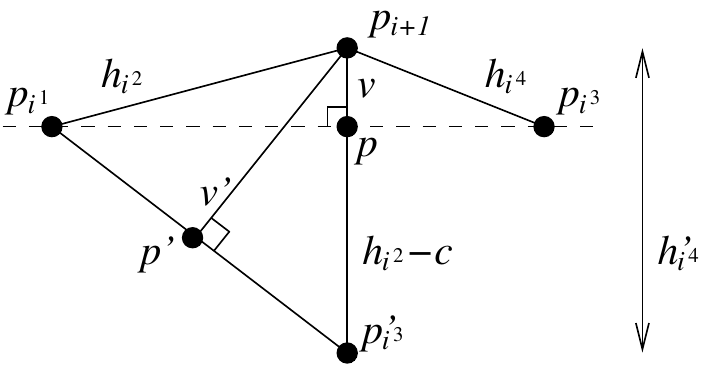}
  \caption{New placement of points given by Algo.~\ref{algo:shift-reference}. $\val p$ (resp. $\val p'$) is the projection of $\val p_{i+1}$ on the line $(\val p_{i^1},\val p_{i^3})$ (resp. $(\val p_{i^1},\val p_{i^3}')$) and $\val v$ (resp. $\val v'$) is the distance from $\val p_{i+1}$ to $p$ (resp. $\val p'$). }
  \label{fig:shift}
\end{figure}

\begin{prop}
\label{prop:shifting}
 Let $\gamma_i$ be the distance to a boundary configuration associated with the instruction 
 $\var p_{i+1}=interCC(\var p_{i^1},\var a_{i^2}, \var p_{i^3},\var a_{i^4})$
 and 
 $(\val p_{i+1},\val p_{i^1},\val a_{i^2}, \val p_{i^3},\val a_{i^4})$
 be values s.t. 
 $\gamma_i(\val p_{i+1}, \val p_{i^1}, \val p_{i^3})\leq \alpha$.
 If $0<\alpha<\frac{1}{2}$ and \ans{$\val p_{i^3}'$} has been obtained with Algo.~\ref{algo:shift-reference} 
 then $\gamma_i(\val p_{i+1}, \val p_{i^1}, \val p_{i^3}')>\alpha$.
\end{prop}

The proof of Prop.~\ref{prop:shifting} is depicted in fig.~\ref{fig:shift},
where $\val p$ is the projection of $\val p_{i+1}$ on the line $(\val p_{i^1},\val p_{i^3})$ and $\val v$ is the distance from $\val p_{i+1}$ to $\val p$.
\ans{Let $\val p_{i^3}'$ be 
the new placement of $\var p_{i^3}$ obtained with Algo.~\ref{algo:shift-reference}.
The criterion}
$\gamma_i$ is computed by considering the distance $\val v'$ from $\val p_{i+1}$ to $\val p'$ where
$\val p'$ is the projection of $\val p_{i+1}$ on the line $(\val p_{i^1},\val p_{i^3}')$.
One has $\frac{\val v'^2}{\val (a_{i^2})^2}=\frac{1}{2}+\frac{\val v}{2\val a_{i^2}}$ and the result follows.
\qed

\section{Correctness and Termination}
\label{section_correctnessTermination}  

Let $\mathcal{S}$ be the homotopy path of~\ref{eq:homofunctionSPM} to which belongs $(\val X^{sk},0)$.
We show here that Algo.~\ref{algo:main-algorithm} terminates, and that
all solutions $(\val X,\val t)$ of~\ref{eq:homofunctionSPM} with $t=1$ lying on $\mathcal{S}$ are in $\mathcal{L}_{sol}$ at the end of Algo.~\ref{algo:main-algorithm}.

\ans{Algo.~\ref{algo:main-algorithm} involves an iterative path tracker that is assumed to track in a finite number of steps a $R$-reduced path if it is a manifold, and if its points are not closer than $\alpha$ to a boundary configuration. The correctness and the termination of our method is proved when considering such an abstract path tracker.
Sec.~\ref{section_results} proposes an implementation of such a path-tracker.
}

Let $R=(I,C_-,\var A_+,\var A_0)$ with $I=(I_i)_{1\leq i \leq l}$ be the RCP given as input of Algo.~\ref{algo:main-algorithm}.
The latter procedure computes a sequence $(R^j)_{j\in N}$ of RCP where $N\subseteq\N_*$, $R^1=R$ and $R^j=(I^j,C_-^j,\var A_+^j,\var A_0^j)$ with  $I^j=(I^j_i)_{1\leq i \leq l}$.
A sequence $(\mathcal{S}^j)_{j\in N}$ of connected pieces of $R^j$-reduced paths is followed, and we note $bj$ the branch of $I^j$ such that $\OnBranch{bj}{R^j}(\mathcal{S}^j)=0$.
We will note $\gamma_{I^j}$ the distance to a boundary configuration associated with $I^j$
and $\OnBranch{bj}{\gamma_{I^j}}(\var A_+^j, \var t)$ for $\gamma_{I^j}(\OnBranch{bj}{I^j}(a(\var t)\uplus\var A_+^j))$.
We will note $\OnBranch{bj}{\varphi_j}$ and $\varphi_j'$ the mappings defined in Eqs.~\ref{eq:varphi} and~\ref{eq:varphip} specialized to the RCP $R^j$.
Let finally $\pi_{\var t}$ be the projection with respect to the $\var t$-coordinate.

The main points of the proof are:
\begin{enumerate}[$(i)$]
 \item $\OnBranch{bj}{\gamma_{I^j}}$ is well defined on $\mathcal{S}^j$,
 \item if $(\val A_+^j, \val t)\in\mathcal{S}^j$ and $\alpha<\frac{1}{2}$ then $\OnBranch{bj}{\gamma_{I^j}}(\val A_+^j, \val t)\geq \alpha$,
 \item $(\mathcal{S}^j)_{j\in N}$ are $1$-dimensional manifolds,
 \item $\bigcup_{j\in N} \OnBranch{bj}{\varphi_j}(\mathcal{S}^j)\subseteq \mathcal{S}$,
 \item $N$ is a finite subset of $\N$ and $\bigcup_{j\in N} \OnBranch{bj}{\varphi_j}(\mathcal{S}^j)=\mathcal{S}$.
\end{enumerate}
Remark that the stopping condition (sc1) is not satisfied on a point $(\val A_+^j, \val t)\in\mathcal{S}^j$, for $j\in N$.
We will prove $(i)$ in Sec.~\ref{subsection_pointi} by showing that \textit{(h4)} and \textit{(h5)} hold for $R^j$, and \textit{(h6)} holds for $R^j$ on $\pi_{\var t}(\mathcal{S}^j)$. Then $(ii)$ holds thanks to Prop.~\ref{prop:shifting}.

$(iii)$ and $(iv)$ are consequences of Prop.~\ref{prop:link}: $\OnBranch{bj}{\varphi_j}$ is a diffeomorphism from $\mathcal{S}^j$ to a connected subset of $\mathcal{S}$.
Notice that $(ii)$ and $(iii)$ are the two conditions under which the abstract path-tracker used in Algo.~\ref{algo:main-algorithm} computes $\mathcal{S}^j$.

$(v)$ is proved in Sec.~\ref{subsection_termination}. It has as a direct consequence that
Algo.~\ref{algo:main-algorithm} terminates and all solutions $(\val X,\val t)$ of~\ref{eq:homofunctionSPM} with $t=1$ lying on $\mathcal{S}$ are found.

\subsection{Proof of Point $(i)$}
\label{subsection_pointi}

As stated in Sec.~\ref{subsubsection_distance}, the distance to a boundary configuration associated with a RCP is well defined when \textit{(h4)}, \textit{(h5)} and \textit{(h6)} hold. It is established in the following proposition, and point $(ii)$ follows as a corollary.

\begin{prop}
\label{prop:assum}
If assumptions \textit{(h4)}, \textit{(h5)} and \textit{(h6)} hold for $R$, 
then $\forall j\in N$, \textit{(h4)} and \textit{(h5)} hold for $R^j$, and \textit{(h6)} holds for $R^j$ at least on $\pi_{\var t}(\mathcal{S}^j)$.
\end{prop}

\noindent\textbf{Proof of Prop.~\ref{prop:assum}:}
Let $j\in N$.
The first instruction of the RCP is never changed in Algo.~\ref{algo:change-RCP}. $I^j_1=I_1$ follows and assumptions \textit{(h4)} holds for $R^j$.

Let $2\leq i \leq l$ and $I^j_i$ be the instruction $\var p_{i+1}=interCC(\var p_{i^1},\var a_{i^2},\var p_{i^3},\var a_{i^4})$.
If $\var a_{i^4}\notin \var A^j_+$ then $I^j_i=I_i$.
Since \textit{(h5)} holds for $R$, \textit{(h5)} holds for $R_j$.

Suppose now $\var a_{i^4}\in \var A^j_+$. 
If $\var a_{i^4}\in \var A_+$, \emph{i.e.} $\var a_{i^4}$ is a driving parameter of $R$, then $I^j_i=I_i$ and $\val a_{i^2}(t)>0$ since \textit{(h6)} holds for $R$ when $\val t\in\PosSup{a}$.

Otherwise, $I^j_i$ is not an original instruction. Let $\var p_{i+1}=interCC(\var p_{i^1},$ $\var a_{i^2}, \var p_{i^3}', \var a_{i^4}')$ be the original instruction $I_i$ and 
\ans{$a_{i^4}$ be the interpolation function for values of $\var a_{i^4}'$}.
$\val a_{i^2}(\val t)$ and $a_{i^4}(\val t)$ does not both vanish according to \textit{(h5)}.
Since condition (sc1) is not satisfied for $(\val A_+^j,\val t)\in\mathcal{S}^j$, $a_{i^2}(\val t)\geq a_{i^4}(\val t)$ holds
and $a_{i^2}(\val t)$ does not vanish on $\pi_{\var t}(\mathcal{S}^j)$.
Thus \textit{(h6)} holds for $R_j$ on $\pi_{\var t}(\mathcal{S}^j)$.
\qed

\begin{cor}
\label{cor:gamma}
If assumptions \textit{(h4)}, \textit{(h5)} and \textit{(h6)} hold for $R$, 
then $\forall j\in N$, $\OnBranch{bj}{\gamma_{I^j}}$ is well defined on $\mathcal{S}^j$.
\end{cor}

\subsection{Proof of Point $(v)$}
\label{subsection_termination}

We consider first the case where $N=\{1,\ldots,n\}$.
The termination condition of step 6 of Algo.~\ref{algo:main-algorithm} is reached, hence the set $\bigcup_{j\in N} \OnBranch{bj}{\varphi_j}(\mathcal{S}^j)$ is diffeomorphic to a circle. 
Since $\bigcup_{j\in N} \OnBranch{bj}{\varphi_j}(\mathcal{S}^j)\subseteq \mathcal{S}$, $\bigcup_{j\in N} \OnBranch{bj}{\varphi_j}(\mathcal{S}^j)=\mathcal{S}$ follows.

We consider now the case $N=\N_*$ and we show that it never happens. 
Algo.~\ref{algo:main-algorithm} constructs a sequence $((\val A_+^j, \val t^j))_{j\in N}$ of points s.t. either $\OnBranch{bj}{\gamma_{I^j}}(\val A_+^j, \val t^j)=\alpha$ or (sc1) is satisfied, and $\OnBranch{bj}{\gamma_{I^j}}(\val A_+^{j-1}, \val t^{j-1})>\alpha$ and (sc1) is not satisfied.
Consider the sequence $((\val X^j, \val t^j))_{j\in N}$ where $(\val X^j, \val t^j) = \OnBranch{bj}{\varphi_j}(\val A_+^j, \val t^j)$.
From point $(iv)$, $((\val X^j, \val t^j))_{j\in N}$ is a sequence of points of $\mathcal{S}$.

From Cor.~\ref{cor:resultSPMadapted}, $\mathcal{S}$ is diffeomorphic to a circle, hence $\mathcal{S}\setminus\{(\val X^{sk}, \val 0)\}$ is diffeomorphic to a bounded open interval, and  
it exists a diffeomorphism $S:]0,1[\rightarrow\mathcal{S}\setminus\{(\val X^{sk}, \val 0)\}$ that maps to $s\in]0,1[$ a point of $\mathcal{S}\setminus\{(\val X^{sk}, \val 0)\}$. 
Reciprocally, $S^{-1}$ maps to a point $(\val X^j, \val t^j)$ a real number $s^j\in]0,1[$.

We show that the sequence $(s^j)_{j\in N}$ satisfies $\forall j\geq 2, s^j>s^{j-1}$ and does not have any accumulation point.
As a consequence, it cannot be infinite.

Suppose it exists $j\geq 2$ s.t. $s^j=s^{j-1}$, hence $(\val X^j, \val t^j)=(\val X^{j-1}, \val t^{j-1})$. 
From Cor.~\ref{cor:gamma}, $\gamma_{I^j}$ is well defined on $\val X^j=\val X^{j-1}$ and 
$\gamma_{I^j}(\val X^{j-1})>\alpha$ hence (sc1) is not satisfied for $(\val X^{j-1}, \val t^{j-1})$, and either 
$\gamma_{I^j}(\val X^{j})=\alpha$ or (sc1) is satisfied for $(\val X^{j}, \val t^{j})$ hence a contradiction follows.
In Algo.~\ref{algo:main-algorithm}, paths $\mathcal{S}^j$ are followed with an orientation that ensures a progression along $\mathcal{S}$, hence we have $s^j>s^{j-1}$.

Suppose now that the sequence $(s^j)_{j\in N}$ has an accumulation point $s^*$, hence $((\val X^j, \val t^j))_{j\in\N}$ has an accumulation point $(\val X^*,\val t^*)$, and it exists a subsequence $((\val X^j, \val t^j))_{j\in N^*}$, with $N^*\subseteq N$, converging to $(\val X^*,\val t^*)$. 
From assumption \textit{(h3)}, there is an index $j^1$ s.t. $\forall j>j^1, \forall 1\leq i^1,i^2\leq m$ the sign of $a_{i^1}(t^j)-a_{i^2}(t^j)$ does not change.
Hence for $j>j^1$, for each instruction $p_{i+1}=interCC(\var p_{i^1}, \var a_{i^2},\var p_{i^3}, \var a_{i^4})$ of $I^j$ with $i\geq 2$, $\val a_{i^2}^j>\val a_{i^4}^j$ and since assumptions \textit{(h5)} and \textit{(h6)} hold, it exists $r>0$ s.t. $\forall j>j^1, \forall i\geq2, a_{i^2}^j>r$.
Now, for each $\epsilon>0$, it exists an index $j^2>j^1$ s.t. $\forall j>j^2$, points and distances between points vary no more than $\epsilon$ between $\val X^{j}$ and $\val X^{j-1}$.
Remark that since signs of $a_{i^1}(t^j)-a_{i^2}(t^j)$ does not change when $j\geq j^2$ grows, (sc1) is satisfied neither for $\val X^{j}$ nor for $\val X^{j-1}$ and it follows that
$\gamma_{I^j}(\val X^{j-1})>\frac{1}{2}$ (from Prop.~\ref{prop:shifting}) and $\gamma_{I^j}(\ans{\val X^{j}})=\alpha<\frac{1}{2}$ when $j\geq2$.
Taking $\epsilon$ sufficiently small (strictly less than $(\frac{1}{2}-\alpha)r$) leads to a contradiction. 
\qed

\section{Generalization to 3D PDSP}
\label{section_generalization}

3D PDSP fit well to the method depicted in this paper: results of \cite{imbach2014leading} as well as reparameterization approach stay valid. 
Given a PDSP $G$ in a 3D geometric universe, solutions of $G$ up to rigid motions are found by fixing a reference consisting in a point $\var p_1$, a line $\var l_1$ and a plane $\var {pl}_1$. Values $\val p_1, \val l_1, \val {pl}_1$ are fixed s.t. $\val p_1\in \val l_1$ and $\val l_1\in \val {pl}_1$.
A CP of $G$ has the structure:
\begin{enumerate}
 \item[$(I_1)$] $\var p_2 = interSL(\var p_1,\var a_1,\var l_1)$
 \item[$(I_2)$] $\var p_3 = interSSP(\var p_1,\var a_2,\var p_2,\var a_3,\var {pl}_1)$
 \item[$(I_3)$] $\var p_4 = interSSS(\var p_1,\var a_4,\var p_2,\var a_5,\var p_3,\var a_6)$
 \item[]      $\ldots$
 \item[$(I_i)$] $\var p_{i+1} = interSSS(\var p_{i^1},\var a_{i^2},\var p_{i^3},\var a_{i^4},\var p_{i^5},\var a_{i^6})$
 \item[]      $\ldots$
\end{enumerate}
where $interSL$ is a sphere-line intersection, $interSSP$ is the intersection of two spheres and one plane, and $interSSS$ is a three spheres intersection.

If $I_i$ is an $interSSS$ instructions, it is decomposed into the two instructions
\begin{enumerate}
 \item[$(I1_i)$] $\var c_{i+1} = interSS(\var p_{i^1},\var a_{i^2},\var p_{i^3},\var a_{i^4})$
 \item[$(I2_i)$] $\var p_{i+1} = interCS(c(\var c_{i+1}), r(\var c_{i+1}),\var p_{i^5},\var a_{i^6})$
\end{enumerate}
where $\var c_{i+1}$ is a circle, $interSS$ is a sphere-sphere intersection,
$c(\var c_{i+1})$ and $r(\var c_{i+1})$ are respectively the center and the radius of $\var c_{i+1}$,
and $interCS$ is a sphere-circle intersection.

$interSSP$ and $interSS$ can be seen as a 3D extension of an $interCC$ instruction in the 2D case. Hence
when assumptions \textit{(h1)} to \textit{(h6)} hold,
boundary configurations encountered on a homotopy path $\mathcal{S}$ of~\ref{eq:homofunctionSPM} for these instructions are the same than boundary configurations in the 2D case.  
The function that measures the distance to a boundary configuration of such instructions is the natural extension of the one associated with an $interCC$ instruction.
When such instructions are swapped, 
reference points are fixed 
as in the 2D case.

\begin{figure}
\centering
  \includegraphics[width=4.5cm]{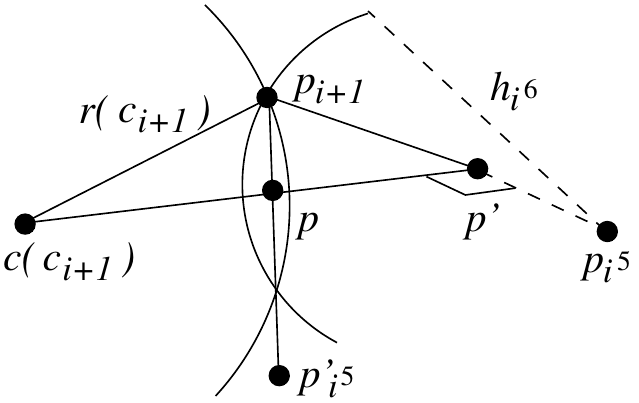}
  \caption{Placement of point $\var p_{i^5}'$ when swapping a $interCS$ instruction. Dashed lines does not belong to the plane of circle $c_{i+1}$.}
  \label{fig:change5}
\end{figure}

Consider now an $interCS$ instruction $I2_i$. A boundary configuration is reached when the circle and the sphere are tangent. The radius of the center is null only for boundary configurations of $I1_i$.
We associate with $(I2_i)$ the function $\gamma_i$ defined as 
\ans{$\gamma_i(\var X) = \dfrac{\var p_{i+1} \var p } {max(\var p_{i+1}c(\var c_{i+1}),\var p_{i+1}\var p_{i^5}) }$ }
where $\var p'$ is the projection of $\var p_{i^5}$ on the plane to which belongs $\var c_{i+1}$, and $\var p$ is the projection of $\var p_{i+1}$ on the line passing by $c(\var c_{i+1})$ and $\var p'$. 
$\gamma_i$ is defined on figures that are not a boundary configuration of $I1_i$. 
If $\var a_{i^6}$ is not already a driving parameter, $\var p_{i+1} = interCS(c(\var c_{i+1}), r(\var c_{i+1}),\var p_{i^5},\var a_{i^6})$ can be swapped with $\var p_{i+1} = interCS(c(\var c_{i+1}), r(\var c_{i+1}),\var p_{i^5}',\var a_{i^6}')$, and the value $\val p_{i^5}'$ for the new reference point $\var p_{i^5}'$ is set as $\val p_{i^5}'=\val p_{i+1} + m\frac{\overrightarrow{\val p_{i+1} \val p}}{\val p_{i+1}\val p}$ where $m$ is the greatest value between $r(\val c_{i+1})$ and $\val a_{i^6}$, and $\val a_{i^6}'$ is set as $m$, as illustrated in Fig~\ref{fig:change5}.
It is then easy to state a proposition equivalent to Prop.~\ref{prop:shifting} for this placement of point.

\section{Implementation and results}
\label{section_results}

Our method has been implemented in \texttt{C++},
giving rise to a program that accepts a PDSP $G$, a RCP, a sketch and provides solutions of $G$.
The path-tracking is achieved by a 
prediction-correction method with 
\ans{an adaptive prediction}
step that is described in Sec.~\ref{subsection_trackingImp}.
It requires to compute partial derivatives of the function $\OnBranch{b}{H}_R$, which appears to be one of the most time consuming step of our method. We propose in Sec. \ref{subsection_differentition} to exploit the acyclic nature of a CP to optimize this operation.
Numerical results related to the solving of four PDSP in 2D and 3D are given in Subsec.~\ref{subsection_results}.
It confirms the efficiency of the approach of \cite{imbach2014leading} to provide several solutions (sometimes all of them) of problems that resist to divide and conquer methods and are too large to be solved by classical numeric solvers providing all the solutions.
Using a RCP as proposed here brings an important speed-up of this approach.

\subsection{Path tracking}
  \label{subsection_trackingImp}
\ans{Homotopy} and 
$R$-reduced paths are followed thanks to
a classical prediction-correction method: prediction is performed along the tangent of the path by an Euler predictor with 
\ans{step $\delta\in[\delta_{min},\delta_{max}]$} and correction by Newton-Raphson iterations.
\ans{The step $\delta$ is doubled (respectively halved) if $2\delta\leq\delta_{max}$ (resp. $\delta\geq2\delta_{min}$) and if the previous correction step did succeed (resp. fail).}
The Jacobian matrices that are required both in prediction and correction steps are numerically computed with finite differences.

In Algo.~\ref{algo:main-algorithm}, when entering for the first time in the main while loop, an orientation (\emph{i.e.} one of the two unit vectors of the tangent) to follow the first $R$-reduced path is arbitrarily chosen.
When entering in the while loop after the RCP has been changed for the $j+1$-th time, the orientation has to ensure the progression along $\mathcal{S}$.
To determine the appropriated orientation,
the last unit vector of the tangent used to track $\mathcal{S}^{j}$ is ``translated'' in the new space where $\mathcal{S}^{j+1}$ is tracked with the application $\varphi_{j+1}'\circ\OnBranch{bj}{\varphi}_{j}$, with notations of Sec.~\ref{section_correctnessTermination}.

Notice that this simple path tracking algorithm does not avoid jumps between paths. Approaches using interval arithmetic (see \cite{kearfott1994interval,faudot2007new,martin2013certified}) could be used to certify the path tracking.
Here we suppose that 
\ans{$\delta_{max}$} is
small enough to follow considered curves 
\ans{while avoiding such jumps.}

\subsection{Differentiation of the CP}
\label{subsection_differentition}
When tracking a $R$-reduced path,
most of the computation time is spent in the evaluation of the underlying RCP.
Most evaluations intervene in the computation of Jacobian matrices by finite differences that needs about $d$ evaluations, where $d$ is the number of driving parameters.
Such matrices are computed at each prediction step and at each iteration of the Newton-Raphson method in a correction step.
Here we exploit the acyclic computation scheme of a RCP to improve its evaluation.

Suppose driving parameters $\var A_+=(\var k_1,\ldots,\var k_d)$ of $R$ appear in instructions $I_{i^1},\ldots,I_{i^{d}}$ with $i^{d}\geq\ldots\geq i^1$. 
When computing with finite differences the derivative with respect to $\var k_j$ of the numerical function associated to $R$,
the geometric objects of the figures resulting of the two evaluations differ only if they are produced by instructions $I_i$ with $i\geq i^{j}$
since $I_{i^j}$ is the first step involving $\var k_j$.

Hence a manner of optimizing the differentiation of a RCP is to evaluate it entirely a first time 
and then to compute the partial derivatives with respect to $\var k_j$ by evaluating the RCP from the step $i^{j}$, for $j$ from $d$ to 1. Our implementation incorporates this optimization.

\subsection{Results}
  \label{subsection_results}
  
We give here numerical results concerning the solving of four PDSP, one in 2D and three in 3D. 
The method depicted here consists in computing the path $\mathcal{S}$ of~\ref{eq:homofunctionSPM} to which belongs the sketch by using a RCP to track $\mathcal{S}$ in the space of driving parameters instead of tracking it in the space of all coordinates.
These two approaches (with and without RCP) 
yield the same number of solutions. 
However, using a RCP brings an important gain in term of running times as it appears in our experiments.
For each problem we also give the
running time and the number of 
solutions obtained when solving the system~\ref{eq:SEParams} with a classical homotopy method
implemented by the free software HOM4PS-2.0 (see \cite{Lee08}).
We did chose homotopy solving as a witness method because as far as we know, it is the sole approach allowing to find all the solutions of large undecomposable problems.
We did choose HOM4PS-2.0 to implement it because among other free softwares implementing homotopy, it seems to be faster to solve sparse systems of polynomials.

\ans{Be given a real solution of a PDSP, the elements of its orbit by the action of the group of reflections through $x$ and $y$ axis (resp. $(x,y)$, $(y,z)$ and $(x,z)$ planes in the 3D case) are also solutions of the PDSP.
In our experiments, the solutions found on $\mathcal{S}$ belong to different orbits.}

\subsubsection{Problems \ans{and parameters settings}}

\newlength{\ind}

\begin{figure}
\begin{minipage}{0.55\linewidth}
\centering
  \includegraphics[width=3.5cm]{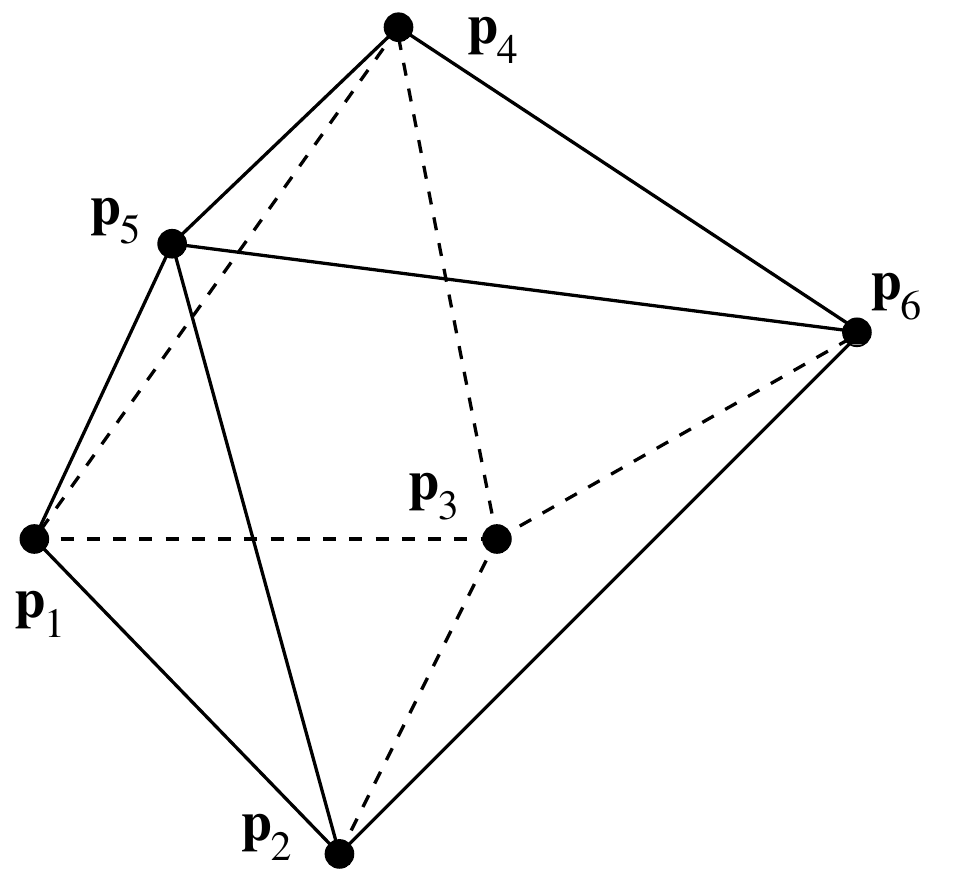}
\end{minipage}
\begin{minipage}{0.4\linewidth}
\centering
  \includegraphics[width=3.5cm]{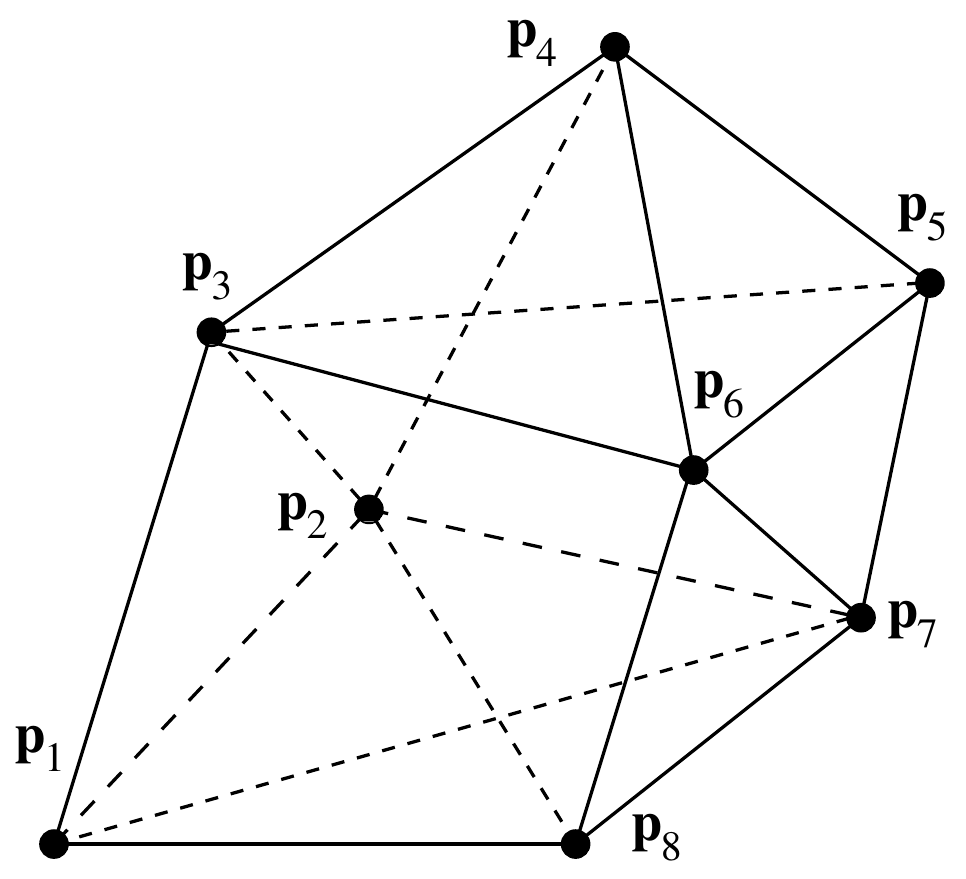}
\end{minipage}
  \caption{
Octahedron (left) and disulfide (right) problems. Edges are distance constraints.
}
  \label{fig:graphes12}
\end{figure}

The goal of the octahedron problem is to construct a solid with $6$ vertices, $12$ edges and $8$ triangular faces knowing the lengths of its $12$ edges
\ans{(see Fig.~\ref{fig:graphes12}).}
This problem \ans{is used in \cite{durand2000systematic} and} is related to the parallel robot called Gough-Stewart platform. It results in a system~\ref{eq:SEParams} of $12$ equations with $12$ unknowns.

The second problem comes from molecular chemistry and is picked up from \cite{porta2007complete}. Coordinates of $8$ points in the 3D space have to be found knowing $18$ distances. 
It corresponds to a disulfide molecule (see right part of Fig. \ref{fig:graphes12}). A valuation  of parameters  is exhibited in \cite{porta2007complete} that leads to $18$ solutions \ans{up to reflexions} all found by a bisection method in more than 10 minutes in \cite{porta2007complete}. 

\begin{figure}
\begin{minipage}{0.55\linewidth}
\centering
  \includegraphics[width=3.5cm]{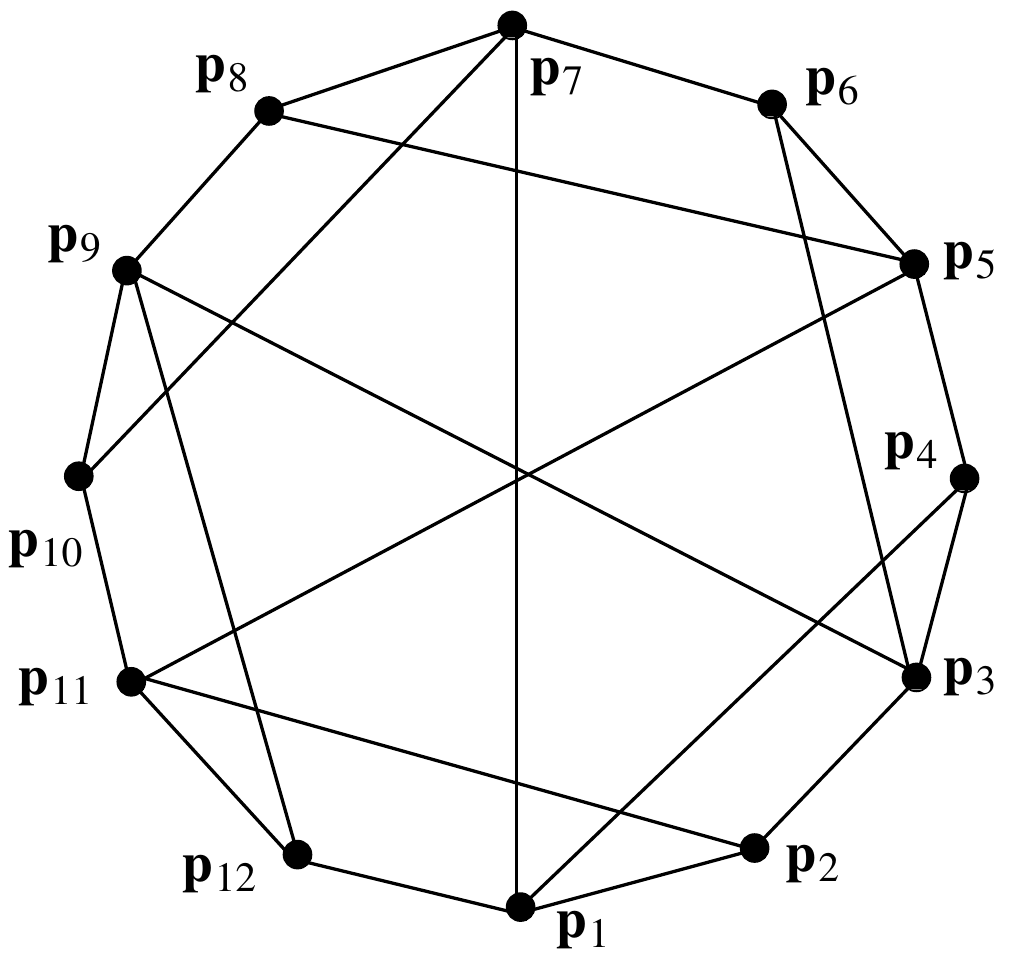}
\end{minipage}
\begin{minipage}{0.4\linewidth}
\centering
  \includegraphics[width=3.5cm]{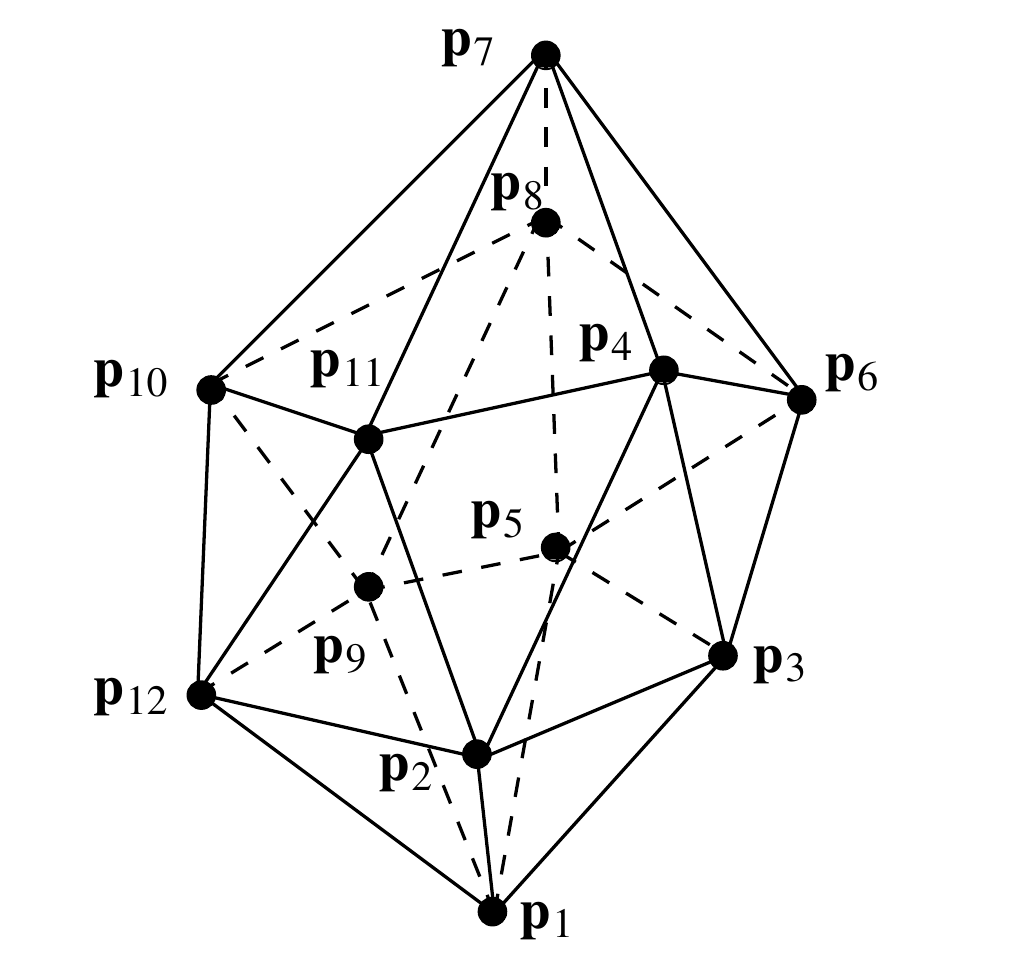}
\end{minipage}
  \caption{
Dodecagon (left) and icosahedron (right) problems. Edges are distance constraints.
}
  \label{fig:graphes34}
\end{figure}

Dodecagon and Icosahedron problems are illustrated on Fig. \ref{fig:graphes34}. 
The former gives rise to a system~\ref{eq:SEParams} with $21$ equations. 
The system~\ref{eq:SEParams} associated to the icosahedron problem involves $30$ equations.

\ans{Values $\val X^{sk}$ of the sketches and $\val A^{so}$ of parameters are given in appendix~\ref{app}}.

\ans{
Interpolation functions have been chosen such that 
$a_m(t)=-2t^2 + (a_m^{so} - a_m^{sk} + 2)t + a_m^{sk}$
and $a_i(t) = (1-t)a_i^{sk} + ta_i^{so}$, for $1\leq i <m$.
The value for $\alpha$ has been set for each problem to $0.1$, what seems to fit well to our algorithm.
The prediction steps vary in the interval $[1^{-10},\delta_{max}]$. As stated above, $\delta_{max}$ has to be chosen small enough to avoid jumping between different paths. The values $\delta_{max}=0.1$ when tracking $\mathcal{S}$ without RCP and $\delta_{max}=0.05$ when using a RCP have been chosen after several trials. Further details and discussions concerning the prediction steps are given in~\ref{subsubsection_details}.
}

\subsubsection{Data of Table~\ref{tab:timetable}}

\setlength{\ind}{0pt}
\begin{table}
 \caption{Sequential running times on an Intel(R) Core(TM) i7-5600U CPU @ 2.60GHz.}

 \begin{center}
  \begin{small}
\ans{
\begin{tabular}{l|c|c|c|c}
                                         & Octahedron     &Disulfide & Dodecagon            & Icosahedron\\[-\ind] 
\hfill{}$m$                              & 12             & 18       & 21                   & 30         \\\hline
\ans{Number of solutions}                &                &          &                      &            \\
\hfill{} \ans{complex}                   & \ans{72}       & \ans{256}& \ans{12580}          & \ans{-}    \\[-\ind]
\hfill{} \ans{real (up to reflections)}  & 4              & 18       & 2                    & -          \\[-\ind]
\hfill{} on $\mathcal{S}$                & 4              & 8        & 2                    & 32         \\\hline
Running times                            &                &          &                      &            \\
\hfill{} HOM4PS-2.0                      & 19.8s          & 3776s    & 10h                  & -          \\[-\ind]
\hfill{} tracking $\mathcal{S}$          &  0.06s         & 0.8s     & 0.07s                & 8.6s       \\[-\ind]
\hfill{} Algo.~\ref{algo:main-algorithm} &  0.02s         & 0.09s    & 0.01s                & 1.5s       \\   
   \end{tabular}
}   
  \end{small}
 \end{center}
\label{tab:timetable}
\end{table}

Table~\ref{tab:timetable} gives for each problem the number $m$ of equations of the system~\ref{eq:SEParams}.
\ans{In group of lines ``Number of solutions'', it first gives the total number of complex solutions of~\ref{eq:SEParams}, all found by HOM4PS-2.0. (line ``complex''). 
The line ``real (up to reflections)'' gives the number of real solutions up to reflections.
The line ``on $\mathcal{S}$'' gives the number of real solutions lying on the path $\mathcal{S}$ to which belongs the sketch, that is obtained with our method. As remarked above, these solutions are different up to reflections. } 

The group of lines ``Running times'' refers to times required to solve each problem with each approach.
\ans{When using HOM4PS-2.0., most efforts are spent to follow paths leading to complex solutions, what explains the large running times in the line ``HOM4PS-2.0.''.}
The line ``tracking $\mathcal{S}$'' refers to the time required to track $\mathcal{S}$ in the space $\R^m\times\R$, without using a RCP.
The line ``Algo.~\ref{algo:main-algorithm}'' refers to the time required to compute solutions on $\mathcal{S}$ with on-the-fly change of RCP; it allows an important gain in term of computation cost.

For the icosahedron problem, we do not give total number of 
solutions and running times for HOM4PS-2.0 since 
\ans{the solving process did not finish}.

\subsubsection{Details on execution}
\label{subsubsection_details}

\ans{Table~\ref{tab:executionDetails} gives for each problem details about paths tracking without RCP (in the first columns) and with RCP (in the second columns). The row ``time $t$ in s'' recalls the execution time in seconds.}

\ans{The row ``smallest $\delta$'' gives the smallest prediction step used during the tracking process.
It shows that tracking a path with a RCP requires to take smaller prediction steps than without a RCP.
Together with the fact that $\delta_{max}$ has to be smaller when using a RCP, it
suggests that $R$-reduced paths have higher curvature than the corresponding paths in the space of figures, and are more difficult to track.}

\ans{The lines ``nb. $i$ of iterations'' and ``$t/i$ in ms'' give the number of iterations of prediction-correction and the average time needed for each iteration. The latter information underlines the main advantage of using a RCP: each iteration of prediction-correction involves less computation than without RCP since the path is tracked in a space of smaller dimension. The line ``time $t$ in s'' shows that this gain counterbalances the drawback mentioned above.}

\ans{The line ``nb. of RCP changing'' gives the number of times the RCP has been changed during the tracking process. This number can be large (see the case of the icosahedron problem): 
as mentioned in the penultimate paragraph of Sec.~\ref{subsubsection_naive}, figures with a boundary configuration are the solutions of systems of $m+1$ equations in $m+1$ unknowns and can be in an exponential number. Hence the approach proposed in this paper could require, in the worst case, to change exponentially many times the RCP.
The rows ``average nb. of DP'' and ``max. nb. of DP'' give respectively the average and the maximum number of driving parameters involved in the RCP and show that the number of driving parameters involved in successive RCP stays much lower than $m$ what keeps the method efficient.
}

\setlength{\ind}{0pt}
\begin{table}
 \caption{Details about path-tracking. For each problem, the first (resp. second) columns refers to the path tracking without (resp. with) RCP.}
 
 \begin{center}
  \begin{small}
\ans{  
   \begin{tabular}{l|cc|cc|cc|cc}
×                                        & \multicolumn{2}{c|}{Octahedron}& \multicolumn{2}{c|}{Disulfide}& \multicolumn{2}{c|}{Dodecagon}& \multicolumn{2}{c}{Icosahedron}\\  [-\ind]
\hfill{}$m$                              & \multicolumn{2}{c|}{12        }& \multicolumn{2}{c|}{18       }& \multicolumn{2}{c|}{21       }& \multicolumn{2}{c}{30         }\\\hline
\hfill{} time $t$ in s                   & 0.06           & 0.02          & 0.8       &  0.09             & 0.07        &  0.01           & 8.6         &   1.5         \\  [-\ind]
\hfill{} \ans{smallest $\delta$}          & \ans{0.1}      & \ans{0.03}    & \ans{0.1} &  \ans{0.01}       & \ans{0.05}  &  \ans{0.02}     & \ans{0.05}  &   \ans{0.001}  \\  [-\ind]
\hfill{} nb. $i$ of iterations           & 157            & 173           & 1097      &  694              & 42          &  46             & 4074        &   3377        \\  [-\ind]
\hfill{} $t/i$ in ms                     & 0.4            & 0.1           & 0.78      &  0.13             & 1.78        &  0.3            & 2.1         &   0.43         \\  \hline
\hfill{} nb. of RCP changing             &      -         & 13            &   -       & 28                & -           & 2               & -           & 250           \\  [-\ind]
\hfill{} average nb. of DP               &      -         & 1.3           &   -       & 1.98              & -           & 4               & -           & 3.8          \\  [-\ind]
\hfill{} max. nb. of DP                  &      -         & 2             &   -       & 3                 & -           & 4               & -           & 6             \\  

   \end{tabular}
}   
  \end{small}
 \end{center}
\label{tab:executionDetails}
\end{table}

\section{Conclusion}
\label{section_conclusion}  

Well-constrained point distance solving problems often have many solutions. 
The existing solvers that offer all the solutions are of limited practical interest because either the class of problems they solve is reduced or their complexity is exponential. But even if not all solutions are needed, several ones similar in shape to the sketch must be provided.

An approach to fulfill this requirement is to use the sketch to define a real homotopy such that
the homotopy path to which belongs the sketch is diffeomorphic to a circle and contains several solutions, that are similar to the sketch in the sense that they belongs to the same homotopy path.

In this article we made this approach
more efficient by reducing the dimension of the space where the homotopy path is tracked by using a symbolic geometric constructions program.
The latter is modified on-the-fly in order to stay robust to critical geometric configurations it could induce.    

This original idea has been implemented to prove its soundness.
In the examples discussed solutions are produced 
\ans{quicker}
when a construction program is used.
Moreover the presented experiments show that our approach can provide several solutions to problems that are too large to be solved with numerical solvers searching all the solutions such as homotopy.

Notice finally that our method could be extended to more general geometric constraints such as angles, collinearities, coplanarities, and so on.
When considering these constraints, homotopy paths are not necessarily diffeomorphic to circles but can converge to special geometric configurations that can be detected when using a construction program to stop the path tracking process.

%




\appendix
\ans{
\section{Numerical values of $\val A^{so}$ and $\val X^{sk}$}
\label{app}
\paragraph{Octahedron}
Values $\val A^{so}$ for parameters are picked up from \cite{durand2000systematic}.
\begin{center}
  \begin{scriptsize}
   \begin{tabular}{lll}
$p_1=(0,0,0)$, \hspace{\ind} & $p_2=(1,0,0)$, \hspace{\ind} & $p_3=(0.5,1.3,0)$ \\
$p_4=(1.2,0.5,1.5)$,               & $p_5=(0,1.8,0.9)$,               &$p_6=(-0.2,-0.7,1.3)$           
   \end{tabular}
  \end{scriptsize}
 \end{center} 
\paragraph{Disulfide molecule}
The reader is referred to~\cite{porta2007complete} to get values $\val A^{so}$ of parameters.
\setlength{\ind}{0.0cm}
\begin{center}
  \begin{scriptsize}
   \begin{tabular}{llll}
$p_1=(0,0,0)$, \hspace{\ind} & $p_2=(2.5,5.3,0)$, \hspace{\ind} & $p_3=(3.2,5.8,5.9)$, \hspace{\ind} & $p_4=(1,8.7,8.3)$, \hspace{\ind}\\
$p_5=(4.2,5.6,7.9)$,         & $p_6=(-2.2,1.3,6.3)$,     &$p_7=(5.5,2.5,6.5)$, &$p_8=(5,0,0)$.       
   \end{tabular}
  \end{scriptsize}
 \end{center}
\paragraph{Dodecagone}
In the tables below, $a_{i,j}$ denotes the value of the parameter of $distance(\var p_i, \var p_j)$.
\setlength{\ind}{0cm}
\begin{center}
  \begin{scriptsize}
   \begin{tabular}{llllll}
$a_{1,2}=3$,       \hspace{\ind} &$a_{2,3}=1.75$, \hspace{\ind} &$a_{3,4}=1.7$, \hspace{\ind} &$a_{4,5}=2.05$ \hspace{\ind} & $a_{5,6}=1.5$, \hspace{\ind} & $a_{6,7}=1.85$, \\
$a_{7,8}=1.45$,                  &$a_{8,9}=1.35$               &$a_{9,10}=1$,                 &$a_{10,11}=1.4$,             & $a_{11,12}=1$,               &$a_{12,1}=0.6$,  \\
$a_{1,4}=4.4$,                   &$a_{3,6}=5.1$,               &$a_{5,8}=3.9$,                &$a_{7,10}=3.05$,             &$a_{9,12}=3.35$,              &$a_{11,2}=4.4$,  \\
$a_{1,7}=4.45$,                  &$a_{3,9}=6.65$,              &$a_{5,11}=4.65$.        
   \end{tabular}
   \medskip
   
   \begin{tabular}{llll}
$p_1=(0,0)$, \hspace{\ind} & $p_2=(2.9,0)$, \hspace{\ind} & $p_3=(3.4,1.8)$, \hspace{\ind} & $p_4=(2.8,3.4)$, \\
$p_5=(1.3,4.6)$,           & $p_6=(0.1,5.7)$,             & $p_7=(-1.0,4.4)$,              & $p_8=(-2.2,3.6)$,\\
$p_9=(-3.0,2.3)$,          &$p_{10}=(-2.4,1.5)$,             &$p_{11}=(-1.2,0.8)$,               &$p_{12}=(-0.6,0.4)$         
   \end{tabular}
  \end{scriptsize}
 \end{center}
\paragraph{Icosahedron}
\setlength{\ind}{0.1cm}
\begin{center}
  \begin{scriptsize}
   \begin{tabular}{llllll}
$a_{1 , 2 }=2.0 $, \hspace{\ind} &$a_{2 , 4 }=4   $, \hspace{\ind} &$a_{4 , 3 }=4.5 $, \hspace{\ind} &$a_{3 , 6 }=3.9 $, \hspace{\ind} &$a_{6 , 5 }=4.45$, \hspace{\ind} &$a_{5 , 8 }=3.8 $,\\
$a_{8 , 9 }=4.4 $,               &$a_{9 , 10}=3.7 $,               &$a_{10 , 12}=4.35$,              &$a_{12, 11}=3.65$,               &$a_{11, 2 }=4.3 $,               &$a_{11, 7 }=2.5 $,\\
$a_{4 , 7 }=2.6 $,               &$a_{6 , 7 }=2.7 $,               &$a_{8 , 7 }=2.8 $,               &$a_{10, 7 }=2.9 $,               &$a_{11, 4 }=3.0 $,               &$a_{11, 8 }=2.95$,\\
$a_{10 , 8 }=3.05$,              &$a_{8 , 6 }=2.9 $,               &$a_{6 , 4 }=3.1 $,               &$a_{2 , 3 }=3.0 $,               &$a_{3 , 5 }=2.9 $,               &$a_{5 , 9 }=3.1 $,\\
$a_{9 , 12}=2.8 $,               &$a_{12, 2 }=3.2 $,               &$a_{1 , 9 }=2.1 $,               &$a_{1 , 5 }=2.2 $,               &$a_{1 , 3 }=2.3 $,               &$a_{1 , 12}=2.4 $.
\end{tabular}
\medskip
   
   \begin{tabular}{llll}
$p_1=(0,0,0)$, \hspace{\ind} & $p_2=(1,0,0)$, \hspace{\ind} & $p_3=(0.5,1,0)$, \hspace{\ind} & $p_4=(11,1.2,1.3)$, \\
$p_5=(-0.5,0.5,1)$,           & $p_6=(1.7,2.3,1.1)$,             & $p_7=(3,3,3)$,              & $p_8=(0.8,1.6,2.2)$,\\
$p_9=(-0,0,2)$,          &$p_{10}=(1.1,1.3,3.2)$,             &$p_{11}=(1.8,1.2,2.1)$,               &$p_{12}=(0.6,0.2,1.3)$         
   \end{tabular}
  \end{scriptsize}
 \end{center}
 }

\end{document}